\documentclass[showpacs,preprintnumbers,prb,eqsecnum,aps]{revtex4-1}
\usepackage{subfig}
\usepackage{appendix}
\usepackage{natbib}
\newcommand{\be}{\begin{equation}}
\newcommand{\ee}{\end{equation}}
\newcommand{\bea}{\begin{eqnarray}}
\newcommand{\eea}{\end{eqnarray}}
\usepackage[dvips]{graphicx,color}
\usepackage{epstopdf}
\usepackage{amsmath,amssymb,bbold,bm}

\begin{document}

\title{Transverse spectral functions and Dzyaloshinskii-Moriya interactions in XXZ spin chains}

\author{Hamed Karimi and Ian Affleck}
\affiliation{Department of Physics and Astronomy, University of British Columbia,
Vancouver, British Columbia V6T 1Z1, Canada}

\begin{abstract}
Recently much progress has been made in applying field theory methods, first developed to study X-ray edge singularities,
 to interacting one dimensional systems  in order 
to include band curvature effects and study edge singularities at arbitrary momentum.  Finding experimental 
confirmations of this theory remains an open challenge. Here we point out that spin chains with 
uniform Dzyaloshinskii-Moriya (DM) interactions provide an opportunity to test these theories since these 
interactions may be exactly eliminated by a gauge transformation which shifts the momentum.  However, 
this requires an extension of these X-ray edge methods to the transverse spectral function of the 
xxz spin chain in a magnetic field, which we provide. 
\end{abstract}

\maketitle

\section{introduction}
One dimensional (1D) interacting systems exhibit unusual  correlation effects dominated by strong 
quantum fluctuations. Fortunately, an array of powerful theoretical methods exist to study this physics, which is 
finding many experimental realizations. One powerful method is based on bosonization,\cite{Ian89,korepin0,gogolin,giamarchi} leading to the 
Luttinger liquid concept. Traditionally these methods are based on low energy effective field theory 
and only apply to the low energy excitations occurring near certain wave-vectors (such as $q=0$). In the 
case of fermion models they begin by linearizing the dispersion relation near the Fermi energy and ignoring 
irrelevant band curvature effects.  However, in the last few years, these bosonization methods have been significantly 
extended by using techniques first developed to study X-ray edge singularities.\cite{glazman06,rodrigo08,glazman08,glazman09,rodrigo09} This has shown that 
band curvature effects, while formally irrelevant in the renormalization group sense,\cite{rodrigo0} can nonetheless have 
important effects on line-shapes of spectral functions even at low energy. Perhaps even more importantly, 
by combining these techniques with Bethe ansatz methods, it has become possible to make exact 
predictions of critical exponents at arbitrary momentum.\cite{rodrigo08,glazman09} That is, spectral functions are predicted to 
have the form $S(q,\omega )\to A[\omega -\omega_L(q)]^{-\eta}$ near singular energies 
$\omega_L(q)$ for arbitrary $q$ where both $\omega_L(q)$, which is not small, and the exponents 
$\eta (q)$ are determined exactly using the Bethe ansatz. Recently, this approach has been
extended to also obtain the amplitudes of correlation functions, $A$, using the Bethe ansatz.\cite{shashi} This new approach has been 
applied to a number of systems including fermions and bosons moving in the continuum,\cite{glazman06,khodas,glazman08,glazman09}
the fermion spectral function for a tight binding model\cite{rodrigo09} and the longitudinal spectral function 
of the xxz S=1/2 spin chain in a magnetic field\cite{rodrigo08,rodrigo07} with Hamiltonian:
\begin{equation}
\label{xxz}
H=J \sum_{i=1}^N \left( S^x_{i}S^x_{i+1}+S^y_{i}S^y_{i+1}+\Delta S^z_{i}S^z_{i+1}\right) - h \sum S^z_i
\end{equation}
The longitudinal spectral function is:
\be S^{zz}(q,\omega )=\sum_j\int_{-\infty}^\infty dt e^{-iqj+i\omega t}<S^z_j(t)S^z_0(0)>\ee
We set the lattice spacing $a=1$. While this is accessible, at arbitrary $q$, 
to neutron scattering experiments these would normally measure a sum of $S^{zz}$ and the 
transverse spectral functions:
\be S^{ss'}(q,\omega )=\sum_j\int_{-\infty}^\infty dt e^{-iqj+i\omega t}<S^s_j(t)S^{s'}_0(0)>\ee
for $(s,s')=(+,-)$ and $(-,+)$.  For non-zero $h$, $S^{+-}$ and $S^{-+}$ are  different.   Due to the ``Jordan-Wigner string operator'', 
the transverse spectral function is much more difficult to compute. While 
the high neutron fluxes available at the Spallation Neutron Source may eventually make 
such experimental confirmation possible, here we explore another route. 

Quasi 1D magnetic compounds that do 
not have link parity symmetry (reflection about the midpoint of a link) will generally have magnetic Hamiltonians 
containing anti-symmetric Dzyaloshinskii-Moriya\cite{dzyaloshinskii,moriya} interactions:
\be \delta H = \sum_j\vec D_j\cdot (\vec S_j\times \vec S_{j+1}).\label{HDM}\ee
The two standard cases are staggered $\vec D_j=(-1)^j\vec D$ 
and uniform $\vec D_j=\vec D$. Staggered DM interactions are invariant under site-parity 
(reflection about a site) but violate symmetry of translation by one site. On the other hand, 
uniform DM interactions have no parity symmetry whatsoever but respect full translation invariance. 
We restrict our further discussion to the case where $\vec D_j\propto \hat z$ so that 
the DM vector is parallel to the easy (or hard) axis of the symmetric exchange interactions, a situation which is sometimes dictated by symmetry.  
Then it is possible to eliminate the DM interactions exactly by a gauge transformation, yielding the standard xxz model of 
Eq.  (\ref{xxz}) with modified parameters. 
This approach was used to study staggered DM interactions in Ref. (\onlinecite{Ian99,Ian1999}) and 
the theory was applied to a number of real materials. Here we consider the case of uniform DM interactions, $\vec D_j=\hat z D$ 
and choose $D>0$.
In this case the gauge transformation is:
\be 
 \tilde{S}^+_j=e^{-i\alpha j}S^+_j  \ \ \tilde{S}^z_j=S^z_j\label{GT}\ee
 where 
 \be
 \alpha=\tan^{-1}\left(\frac{D}J\right).\label{alpha}\ee
By use of  above unitary transformations we have
\begin{equation}
 H=\mathcal{J} \sum_i \left( \tilde{S}^x_{i}\tilde{S}^x_{i+1}+\tilde{S}^y_{i}\tilde{S}^y_{i+1}+\Delta_{eff} \tilde{S}^z_{i}\tilde{S}^z_{i+1}\right)
-h\sum_i\tilde{S}^z_i.\label{gaugeH}
\end{equation}
where the new exchange coupling and anisotropy parameters are given by 
\be \mathcal{J}=\sqrt{J^2+D^2}  \hspace*{1in} \Delta_{eff}=\Delta \cos(\alpha).\label{Deff}\ee

The electron spin resonance (ESR) adsorption intensity, in standard Faraday configuration, is proportional to 
the transverse spectral function at $q=0$, since the wave-vector of microwave photons is 
 much less than the inverse lattice spacing.  After the gauge transformation, the ESR 
 intensity is therefore proportional to $S^{+-}$ and $S^{-+}$, for the Hamiltonian of Eq.  (\ref{xxz}) at $q=\alpha$. [By using 
 circularly polarized microwave radiation both $S^{+-}(\alpha ,\omega )$ and $S^{-+}(\alpha ,\omega )$ 
 could be measured separately.]  Thus the edge singularities predicted by X-ray edge methods at 
 a non-zero wave-vector $\alpha$ given by Eq.  (\ref{alpha}) are directly measured by ESR. 
 ESR on spin chain compounds with uniform DM interactions therefore would provide 
 a powerful probe of the new bosonization predictions. Quasi-1D spin-1/2 antiferromagnetic insulators containing 
DM interactions with a uniform component 
include Cs$_2$CuCl$_4$\cite{cscuclt,cscucle}  and KCuGaF$_6$.\cite{kcgf} This provides a strong motivation to extend the X-ray edge methods to study edge 
 singularities in the transverse spectral functions of the xxz chain in a magnetic field. 
 
 In the next section we review results on the transverse spectral function using standard bosonization and 
 then show that band curvature effects (in the equivalent fermion model) render these results invalid 
 close to edge singularities. In Sec. III we apply X-ray edge methods to the model
 obtaining new results on the leading edge singularities. In Sec. IV sub-dominant singularities are discussed. Section V discusses ESR with uniform DM interactions, 
 based partly on the results of Sec. III. Sec. VI contains conclusions and open questions.

\section{Spectral function of the xxz spin chain}

The xxz $S=1/2$ model of Eq.  (\ref{xxz}) is equivalent to an interacting spinless fermion model by the Jordan-Wigner transformation:
\begin{eqnarray}
\label{jordan}
 S^z_i & = & c^{\dagger}_i c_i -\frac{1}{2} \nonumber \\
 S^-_j & = & (-1)^j \exp(i \pi \sum_{k<j} c^{\dagger}_k c_k ) c_j
\end{eqnarray}
The Hamiltonian (\ref{xxz}) is transformed to
\begin{eqnarray}
 H & = & -\frac{J}{2}\sum(c^{\dagger}_i c_{i+1} +h.c.) - h \sum c^{\dagger}_i c_i  \nonumber \\
 & + & J \Delta \sum (c^{\dagger}_i c_i -\frac{1}{2})(c^{\dagger}_{i+1} c_{i+1} -\frac{1}{2}).\label{Hferm}
\end{eqnarray}
Note the factor of $(-1)^j$ in the second line of Eq.  (\ref{jordan}), necessary for the first term in the fermionic
Hamiltonian, Eq.  (\ref{Hferm}) to have the standard minus sign. $h$ is the chemical potential of the fermionic 
model with $h=0$ corresponding to half-filling. 
For the non-interacting case $\Delta=0$ the Hamiltonian is just a free fermion model and by going to momentum space 
 \[c_p = \frac{1}{\sqrt{N}} \sum c_j e^{i p j}\]
where $p=2\pi n/N$ for periodic boundary conditions, the energy spectrum of non-interacting fermion model is found to be
 \begin{equation}
  H= \sum (-J \cos(p) -h) c^{\dagger}_p c_p
 \end{equation}
 For this model the longitudinal spectral function $S^{zz}$ is known exactly and the result is
\begin{equation}
\label{szz}
 S^{zz}(\omega,q)= \frac{\theta(\omega-\omega_L(q))\theta(\omega_U(q) - \omega)}{\sqrt{\left(2J \sin \frac{q}{2}\right)^2-\omega^2}}.
\end{equation}
We see that $S^{zz}$ is non-zero only in finite frequency range for a fixed momentum $q$, these lower and upper thresholds for $k_F<\pi/2$ are given by

\begin{eqnarray}
 \omega_U(q) = 2 J\sin \frac{|q|}{2} \sin(k_F + \frac{|q|}{2}) \nonumber \\
 \omega_L(q) = 2 J \sin \frac{|q|}{2} \sin(k_F - \frac{|q|}{2})
\end{eqnarray}
and for $k_F = \pi /2$
\begin{eqnarray}
 \omega_U(q) = 2 J \sin \frac{|q|}{2}   \nonumber \\
 \omega_L(q) =  J \sin |q|
\end{eqnarray}
 From Eq. (\ref{szz}), it is easily seen that the support of $S^{zz}$, for $\Delta =0$, is restricted to the interval $[\omega_L,\omega_U]$. 
The nature of these lower and upper thresholds is  different for zero and non-zero magnetic field, for non-zero magnetic field and $k_F<\pi/2$ the lower threshold is given by creating one deep hole with
 momentum $q$, whereas the upper threshold is achieved by excitation of one high-energy electron with momentum $q$. But for zero magnetic field, $k_F=\pi/2$, particle-hole symmetric case, 
 the lower threshold is achieved either by creating one deep hole or one high energy electron but higher threshold is given by symmetric particle hole excitation
around Fermi point with momentum of each $q/2$. 

In contrast to the longitudinal spectral function, even for $\Delta=0$ and  having  the entire spectrum at hand, finding the transverse spectral function is a very difficult problem.
 The reason for this complication is the complicated 
form of the transverse spin operators, which  have the string operator in their fermionic form Eq.  (\ref{jordan}).
  The equal time transverse correlation function, in this case, is known exactly\cite{mccoy} but less is known about dynamical correlations. The non-local nature of string 
operator is such that we can not even evaluate simply the lower and upper threshold for $S^{-+}$ and $S^{+-}$. By applying $S^+_j$ to the vacuum, the exponential factor of it will allow creation of
 any number of particle hole excitations
 \[S^+_i |0> \: = \: \exp(i \pi \sum_{j<i}c^{\dagger}_jc_j )c^{\dagger}_i |0> \: = \: \sum _n\frac{(i\pi\sum_{j<i}c^{\dagger}_jc_j )^n}{n!}c^{\dagger}_i |0>\]
So it is obvious that potentially $S^{-+}(\omega,q)$ could be non-zero for fixed $q$ and any $\omega$.  However, in general, we expect an infinite number of progressively 
weaker singularities, 
extending down to zero energy, similar to the case of the longitudinal structure function in a non-zero magnetic 
field.\cite{rodrigo09}
After reviewing the standard bosonization results in sub-section A, we 
discuss effects of irrelevant operators arising from band curvature in sub-section B.

\subsection{Bosonization approach}
\label{sec:bosonize}
As mentioned before, calculating the dynamical correlations for xxz model is very difficult, and is one of the most studied problems of one dimensional spin chains.
By using field theory methods known as bosonization\cite{giamarchi,Ian89,gogolin} we could get some information about low-energy effective description of these correlations. Here we review the results of bosonization for the transverse spectral 
function  of the xxz Hamiltonian Eq.  (\ref{xxz}). 
We saw that the xxz model is equivalent to an interacting fermion model by using the Jordan-Wigner transformation. Now by bosonizing the fermion operators we obtain the bosonic representation of spin operators. In the low-energy effective theory, we only include excitation around the two Fermi points so fermionic operators can be written in the following form
\be c_j \approx \psi_R(x) e^{i k_F j}+ \psi_L(x) e^{-i k_F j}\ee
where $\psi_R$ and $\psi_L$ are slowly varying fields. Now by using Abelian bosonization\cite{coleman} we could write the fields as follow
\begin{equation}
\psi_{R,L}(x) \sim \frac{1}{\sqrt{2\pi \epsilon}}e^{-i\sqrt{2\pi} \phi_{R,L}(x)}
\end{equation}
where $\epsilon \approx k^{-1}_F$ is a short-distance cut off, and $\phi_{R,L}$ are the right and left components of bosonic fields $\tilde{\phi}$ and its dual field $\tilde{\theta}$
\begin{eqnarray}
\tilde{\phi} & = & \frac{\phi_{L}-\phi_{R}}{\sqrt{2}} \nonumber \\
\tilde{\theta} & = & \frac{\phi_{L}+\phi_{R}}{\sqrt{2}} \nonumber
\end{eqnarray}
In terms of these bosonic field the effective bosonic description of xxz Hamiltonian is
\begin{equation}
\mathcal{H}_{LL} = \frac{v}{2}\left( K (\partial_x \tilde{\theta})^2 + \frac{1}{K} (\partial_x \tilde{\phi})^2\right)
\end{equation} 
where $v$ and $K$ are spin velocity and Luttinger parameter, respectively. These parameters are known exactly, for zero magnetic field, from Bethe ansatz calculations:\cite{korepin0}
\be v= J{\pi\over 2} {\sqrt{1-\Delta^2}\over \arccos\Delta},\ \  K=[2-2\arccos(\Delta)/\pi]^{-1}.\label{vK}\ee
 $K=1$ for $\Delta =0$ and $K=1/2$ for $\Delta =1$.  For finite magnetic 
field there is no closed expression for them, but they can be evaluated numerically by using Bethe ansatz. The above Hamiltonian is equivalent to the free boson Hamiltonian
with canonical transformation of bosonic fields as
\be \tilde{\phi} =\sqrt{K}\phi ,\ \   \tilde{\theta} = \theta/\sqrt{K}.\label{can}\ee
The Hamiltonian then becomes the standard free boson hamiltonian: 
\begin{equation}
\mathcal{H}_{LL} = \frac{v}{2}\left( (\partial_x \theta)^2 + (\partial_x \phi)^2\right)
\end{equation} 
Now by knowing the Hamiltonian in terms of bosonic field, let's look at the form of spin operators in terms of them. This is quite straightforward for $S^z_j$:
\be S^z_j\approx m+:\psi_L^\dagger\psi_L:+:\psi_R^\dagger \psi_R:+(\psi^\dagger_L\psi_R+h.c.)=m+\sqrt{\frac{K}{\pi}}\partial_x \phi +C^z \cos \left[\sqrt{4\pi K} \phi + (2\pi m + \pi)x\right] 
\label{Szlow}\ee
where $:\ldots :$ denotes normal ordering and $m=<S^z_h>$ is the magnetization, which is related to the Fermi wave-vector by the exact relation:
\be 2k_F=2\pi m+\pi \ee
For weak fields, 
\be m\to Kh/(\pi v).\ee
 $C^z$ is a non-universal constant. 
To obtain the low energy representation of $S^+_j$ from Eq.  (\ref{jordan}) we use:
\be \psi_{R/L}\propto e^{-i\sqrt{\pi /K}\theta \pm i\sqrt{\pi K}\phi}\ee
We also approximate the exponential of the Jordan-Wigner string operators using:
\be i \pi \sum_{l<j} c^{\dagger}_l c_l\approx i\pi \int_{-\infty}^{j}dy(k_F+\sqrt{K/\pi}\partial_y\phi )=\hbox{constant}+i\pi k_Fj +i\sqrt{\pi K} \phi (j).\label{contstring} \ee
Following the standard bosonization approach\cite{Ian89} we have ignored the oscillating term in $c^\dagger_jc_j$ in the exponential 
of the Jordan-Wigner string operator, but we will consider it in the next
sub-section. Note that $\exp [i\sum_{j<i}c^\dagger_jc_j]$ is Hermitian, taking eigenvalues $\pm 1$. On the other hand, the exponential 
of the continuum limit operator in Eq.  (\ref{contstring}) is not Hermitian. To deal with this problem the standard approach\cite{Ian89} is 
to instead take the continuum limit of $\cos [\pi \sum_{l<j}c^\dagger_lc_l]$:
\be \cos [\pi \sum_{l<j}c^\dagger_lc_l]\propto e^{i\pi k_Fj}e^{i\sqrt{\pi K}\phi (j)} + h.c.\ee
Substituting these low energy limit formulas into the second of Eq.  (\ref{jordan}) gives:
\bea S^-_j&\propto& (-1)^je^{-i\sqrt{\pi /K}\theta}\left[e^{ik_Fj}e^{i\sqrt{\pi K}\phi}+e^{-ik_Fj}e^{-i\sqrt{\pi K}\phi}\right]\cdot \left[e^{ik_Fj}e^{i\sqrt{\pi K}\phi}+e^{-ik_Fj}e^{-i\sqrt{\pi K}\phi}\right]
\nonumber \\
&=&e^{-i\sqrt{\pi /K}\theta}\left[C(-1)^j+C^-\cos (2\pi mx+\sqrt{4\pi K}\phi )\right]\label{S-low}
\eea
where $C$ and $C_-$ are two other non-universal constants. 
It can be seen from Eqs. (\ref{Szlow}) and (\ref{S-low})  that the applied magnetic field induces a shift in momenta of both transverse and longitudinal spin operators, but in different ways. For the  longitudinal operator,
 it shifts only the staggered part but for transverse spin it shifts only the uniform part.  This shift is small for weak fields where the approximation of ignoring 
 band curvature is valid.
 
\begin{figure}
  \centering
  \subfloat[$S^{+-}$]{\label{fig:g+}\includegraphics[scale=.3]{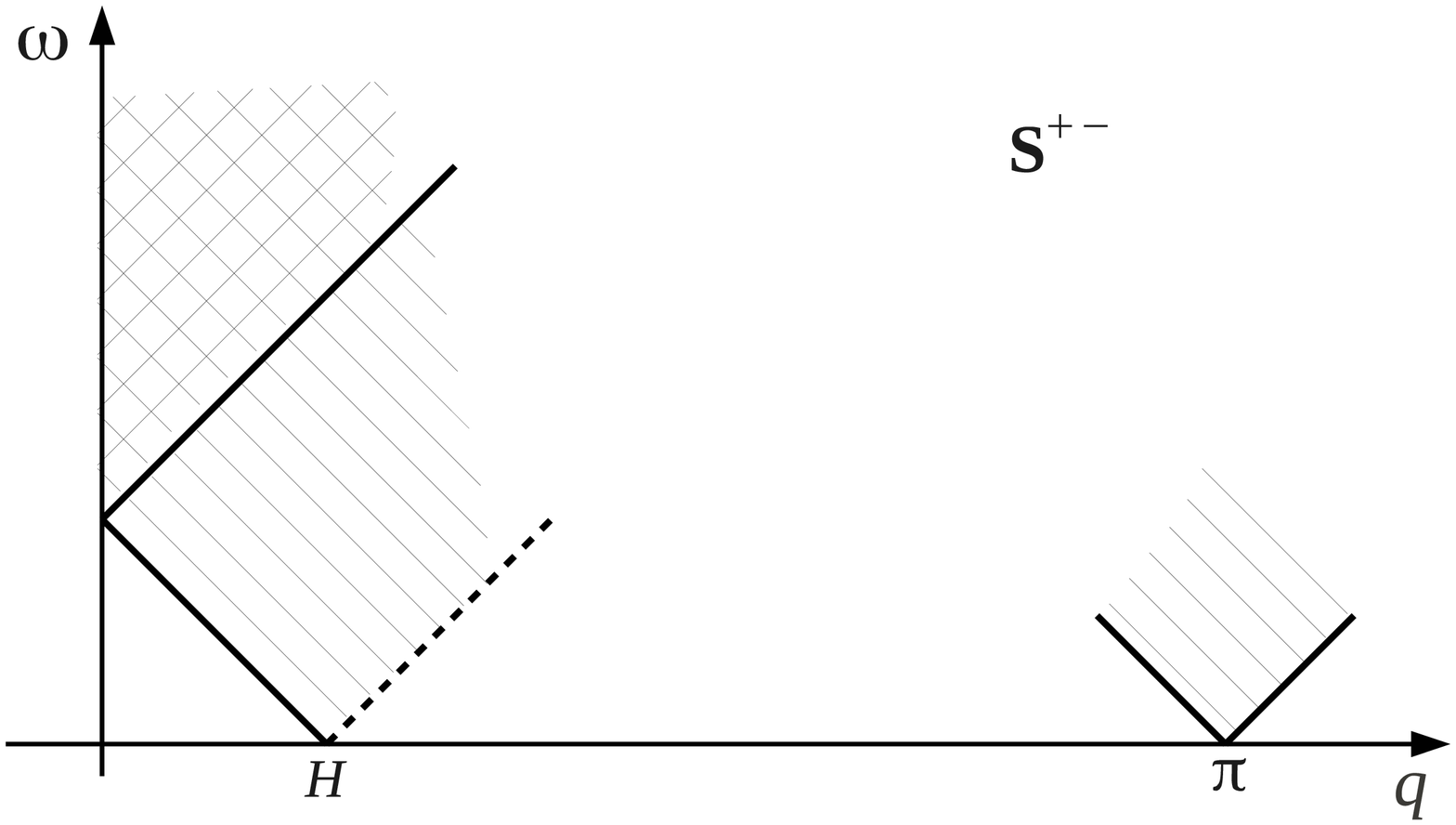}}                
  \subfloat[$S^{-+}$]{\label{fig:g-}\includegraphics[scale=.3]{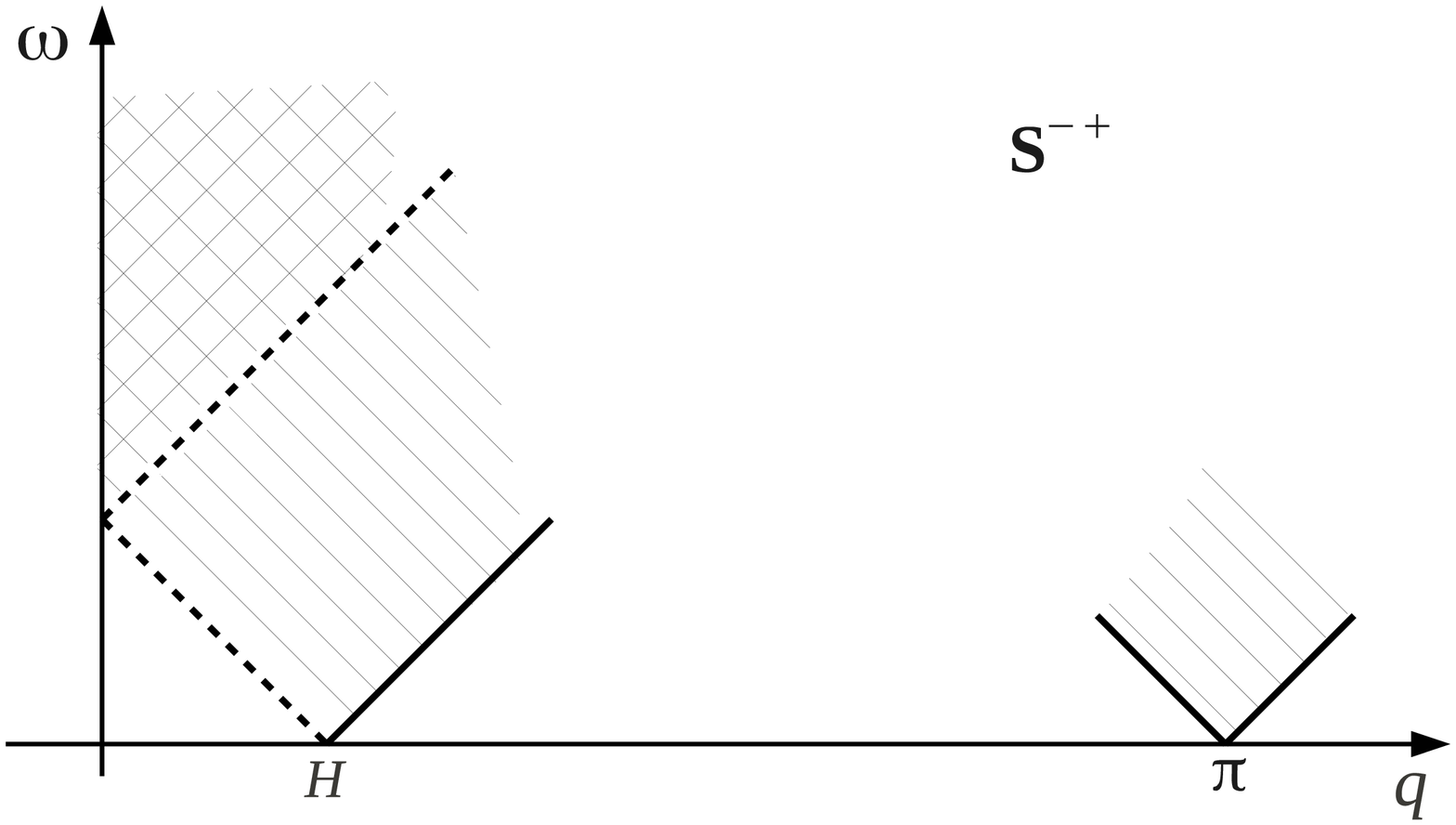}}
  \caption{Singular points of zero temperature transverse spin spectral function of xxz model, predicted by bosonization. The solid lines indicate diverging singularities and dashed line the vanishing singularities.}
  \label{fig:transverse}
\end{figure}

Let us focus on transverse structure function  near $q \approx 0$ for weak fields. At zero temperature we have 
\begin{eqnarray}
<S^{+}_j(t)S^-_0(0)& >  & \propto \frac{e^{-i H x}}{(vt+x-i\epsilon)^{2+2\eta}(vt-x -i\epsilon)^{2\eta}} \nonumber \\
 & + & \frac{e^{i H x}}{(vt-x-i\epsilon)^{2+2\eta}(vt+x -i\epsilon)^{2\eta}} \nonumber \\
\end{eqnarray} 
where $x=j$, $H$ and $\eta$ are given by
\bea
H & = & 2\pi m \nonumber \\
\eta & = & \frac{(1-2K)^2}{8K}\label{eta}
\eea
obeying $\eta <1/8$ for $|\Delta | <1$, and $\epsilon$ is a positive quantity of order the lattice spacing. 
 By taking the Fourier transform, the spectral function $S^{+-}$ is
\begin{eqnarray}
\label{bosonexpo}
S^{+-}(q,\omega) & \propto &  \theta(\omega-v|q+H|)\frac{(\omega+v(q+H))^{1+2\eta}}{(\omega-v(q+H))^{1-2\eta}} \nonumber \\
& +& \theta(\omega-v|q-H|)\frac{(\omega-v(q-H))^{1+2\eta}}{(\omega+v(q-H))^{1-2\eta}} .\nonumber \label{spec}
\end{eqnarray}
Note that the first term has a diverging threshold at $\omega =v(q+H)$ for $q+H>0$ and a vanishing threshold at $\omega =-v(q+H)$ for $q+H<0$.  The 
second term is the parity transform ($q\to -q$) of the first. Diverging and vanishing thresholds are indicated by solid and dotted lines in Fig.  [\ref{fig:g+}]. 
Note that, two diverging thresholds occur in $S^{+-}(q,\omega )$ for  $|q|<H$ which cross each other at $q=0$ but that there is only one diverging thresholds for $|q|>H$. 
 $S^{-+}$ is obtained from $S^{+-}$ by the transformation $H\to -H$.  (Recall that we are assuming $H>0$.)
 
 \begin{equation}
S^{-+}(q,\omega) \propto  \theta(\omega-v|q-H|)\frac{(\omega+v(q-H))^{1+2\eta}}{(\omega-v(q-H))^{1-2\eta}}+\theta(\omega-v|q+H|)\frac{(\omega-v(q+H))^{1+2\eta}}{(\omega+v(q+H))^{1-2\eta}}
\end{equation}
Its diverging and vanishing thresholds are shown in Fig.  [\ref{fig:g-}] Note that, for $S^{-+}$, no diverging thresholds occur for $|q|<H$ and a single diverging threshold occurs for $|q|>H$. 

For ESR applications we will be especially interested in the case $\Delta$ slightly 
less than 1 and small $H$ corresponding to $K$ slightly greater than 1/2 and thus $\eta \ll 1$. Then it is important to note that 
the $\eta$ dependence of the constant factor in $S^{+-}$ and $S^{-+}$ is\cite{schulz}  $\propto \sin^2(2\pi \eta )\Gamma (-1-2\eta )\Gamma (1-2\eta )$ which vanishes 
linearly with $\eta$. Here $\Gamma$ is Euler's Gamma function.  To study the line shape at $H=0$ and $\eta \to 0$, we take into account that this expression for $S(q,\omega )$ is only 
valid for a finite range of $\omega$, $v|q|<\omega <\Lambda$ for an upper cut off $\Lambda$, of order $J$ or less. 
We then use the fact that 
\begin{equation}
\lim_{\eta \rightarrow 0}  \int^{\Lambda}_{vq}  \frac{2\eta \: \theta(\omega - vq)}{(\omega -vq)^{1-2\eta}} = \lim_{\eta \rightarrow 0} \Lambda^{2\eta} = 1.
\end{equation}
Therefore we have 
\[\lim_{\eta \rightarrow 0}\frac{2\eta \: \theta(\omega - vq)}{(\omega -vq)^{1-2\eta}} = \delta (\omega - vq )\]
and thus the term with a diverging threshold approaches
\begin{equation}
S(q,\omega) \propto v|q| \delta (\omega -vq).\ \  (\eta \to 0).
\end{equation}

 \begin{figure}
  \centering
  \subfloat[$h=0$]{\label{fig:g}\includegraphics[scale=.7]{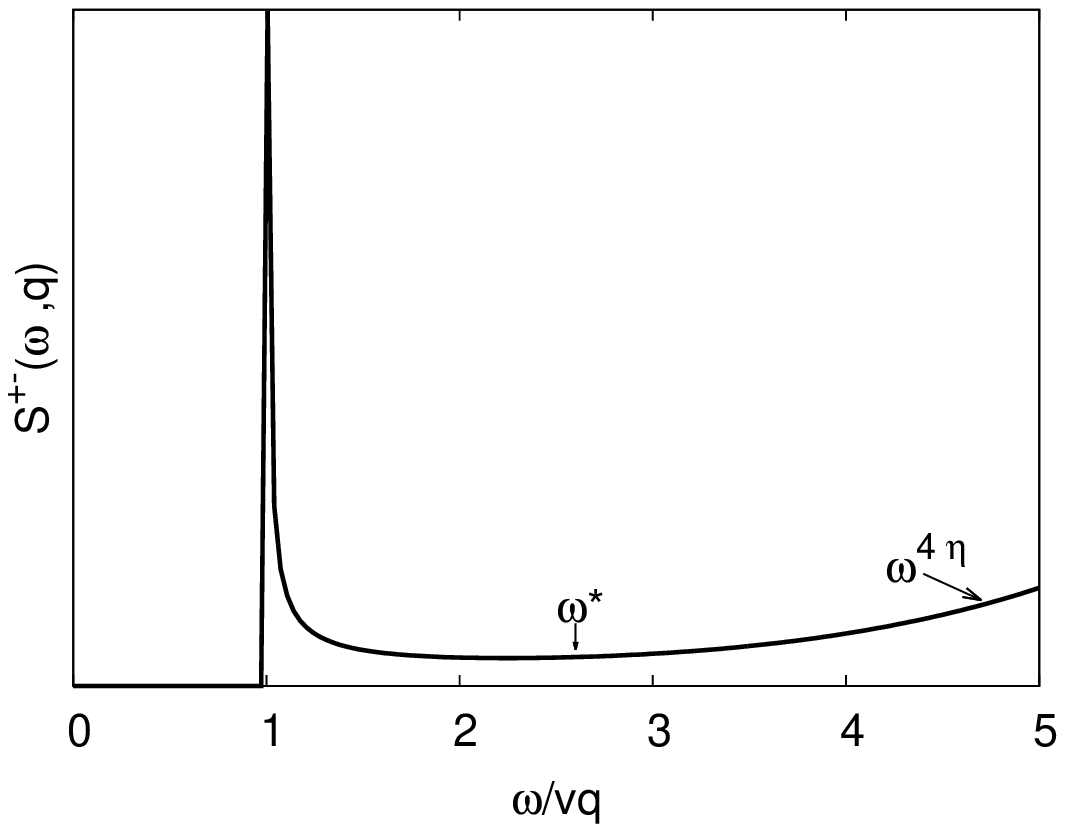}}                
  \subfloat[$h \neq 0$]{\label{fig:g11}\includegraphics[scale=.7]{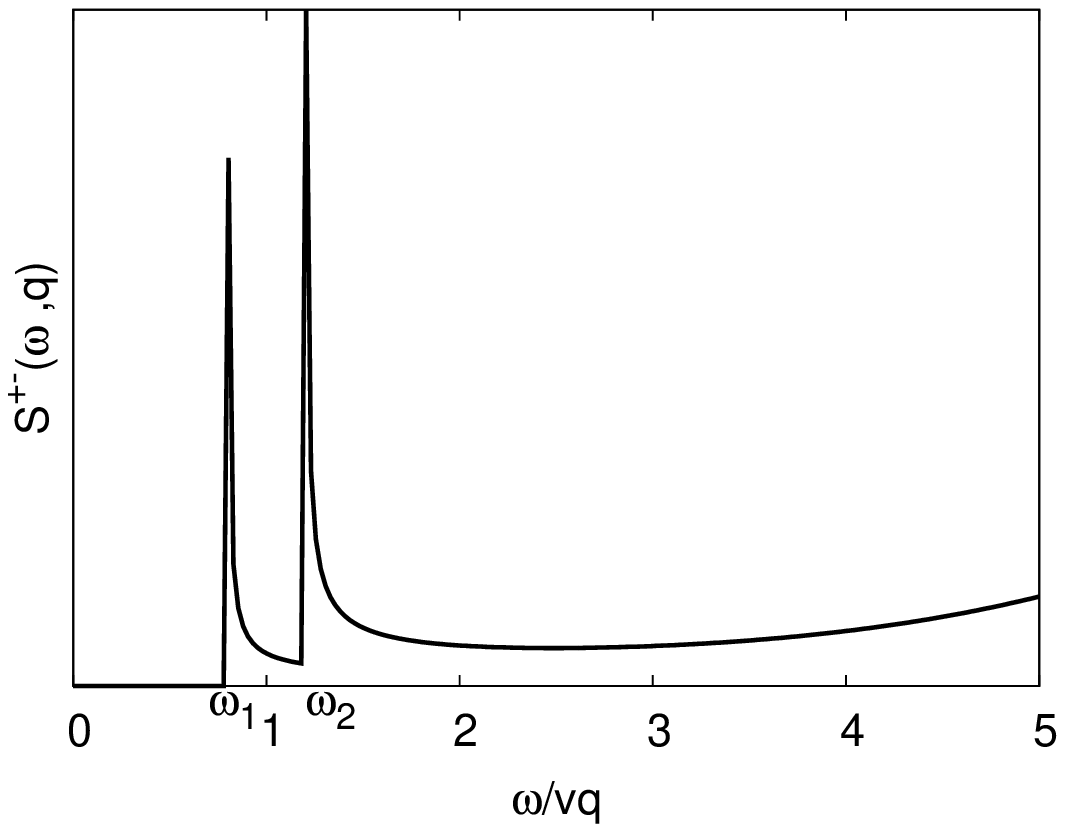}}
  \caption{Zero temperature transverse spectral function $S^{+-}(\omega,q)$ predicted by bosonization for fixed $q$. a) shows transverse spectral function for zero magnetic field and b) is for non-zero magnetic field with $q<H$.}
  \label{fig1}
\end{figure}


For a fixed momentum $q$ and small $\eta$, $S(q,\omega )$ as function of $\omega$ is depicted in Fig. [\ref{fig1}].  Here we show only one term with a 
diverging threshold.  It is zero for frequencies such that $\omega < v q$, and it has a local 
 minima at point $\omega^*$  given by 
\begin{equation}
 \label{min}
\omega^*=\sqrt{\frac{1-\eta}{\eta}}v q \approx \frac{v q}{\sqrt{\eta}}
\end{equation}
So for $\eta \approx 0.1$ we get $\omega^* \approx 3 v q$.  We should also be careful about the cases of very small anisotropy; from Eq (\ref{min}) we see that for fixed momentum $q$, as anisotropy gets smaller and smaller $\omega^*$ becomes larger and larger, so it seems that we are getting out of the region where bosonization works. The results of bosonization  are reliable 
below  some cutoff $\Lambda$; then the consistency relation $\omega^*<\Lambda$ will gives us a restriction on momentum $q$ such that we must have $vq<\sqrt{\eta} \Lambda$. 

\begin{figure}[h!]
  \centering
  \subfloat[$|q|\lesssim H$ for $S^{+-}$]{\label{fig:gull}\includegraphics[scale=.3]{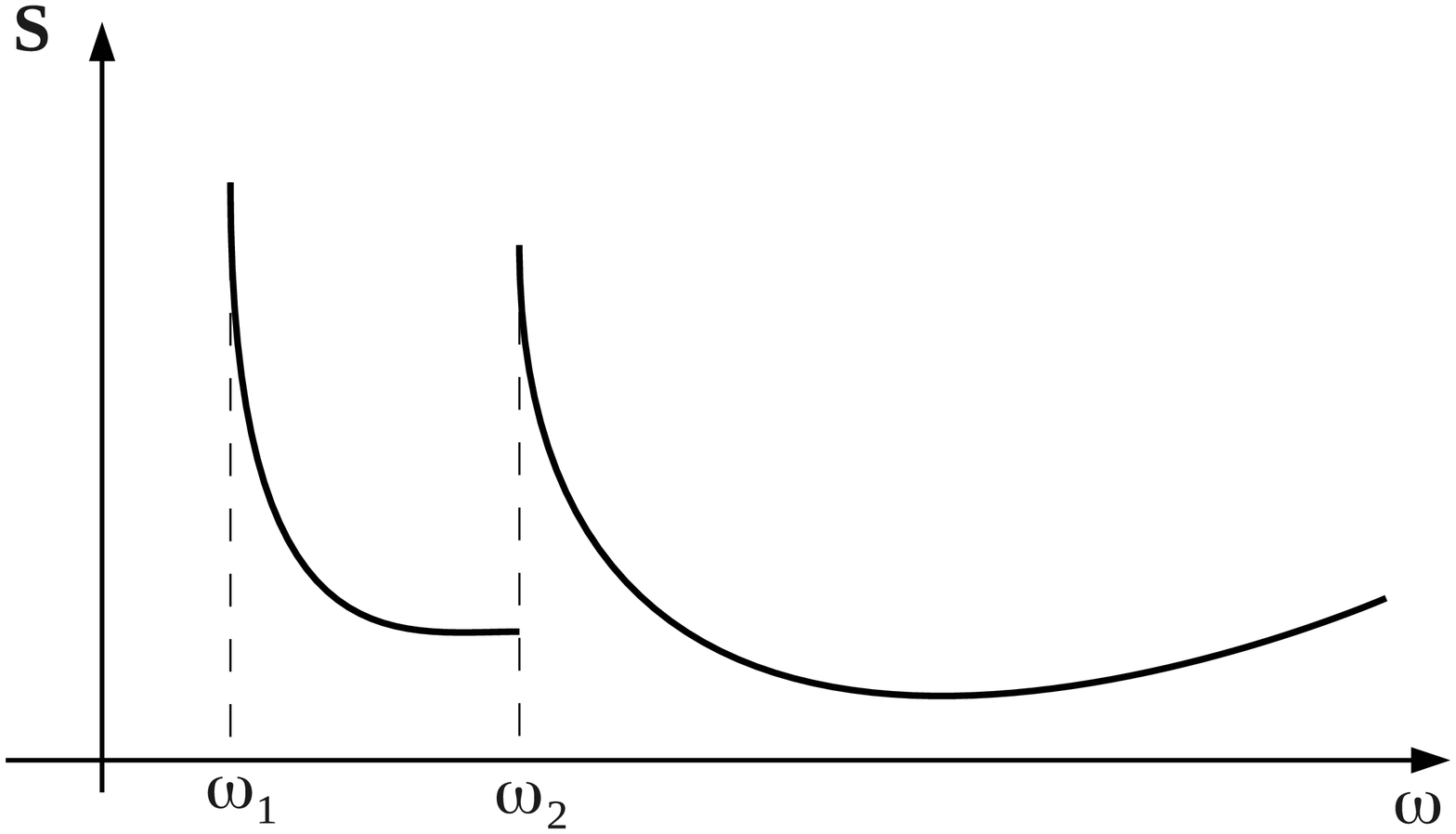}}                
  \subfloat[$H\lesssim |q|$ for $S^{+-}$]{\label{fig:tiger}\includegraphics[scale=.3]{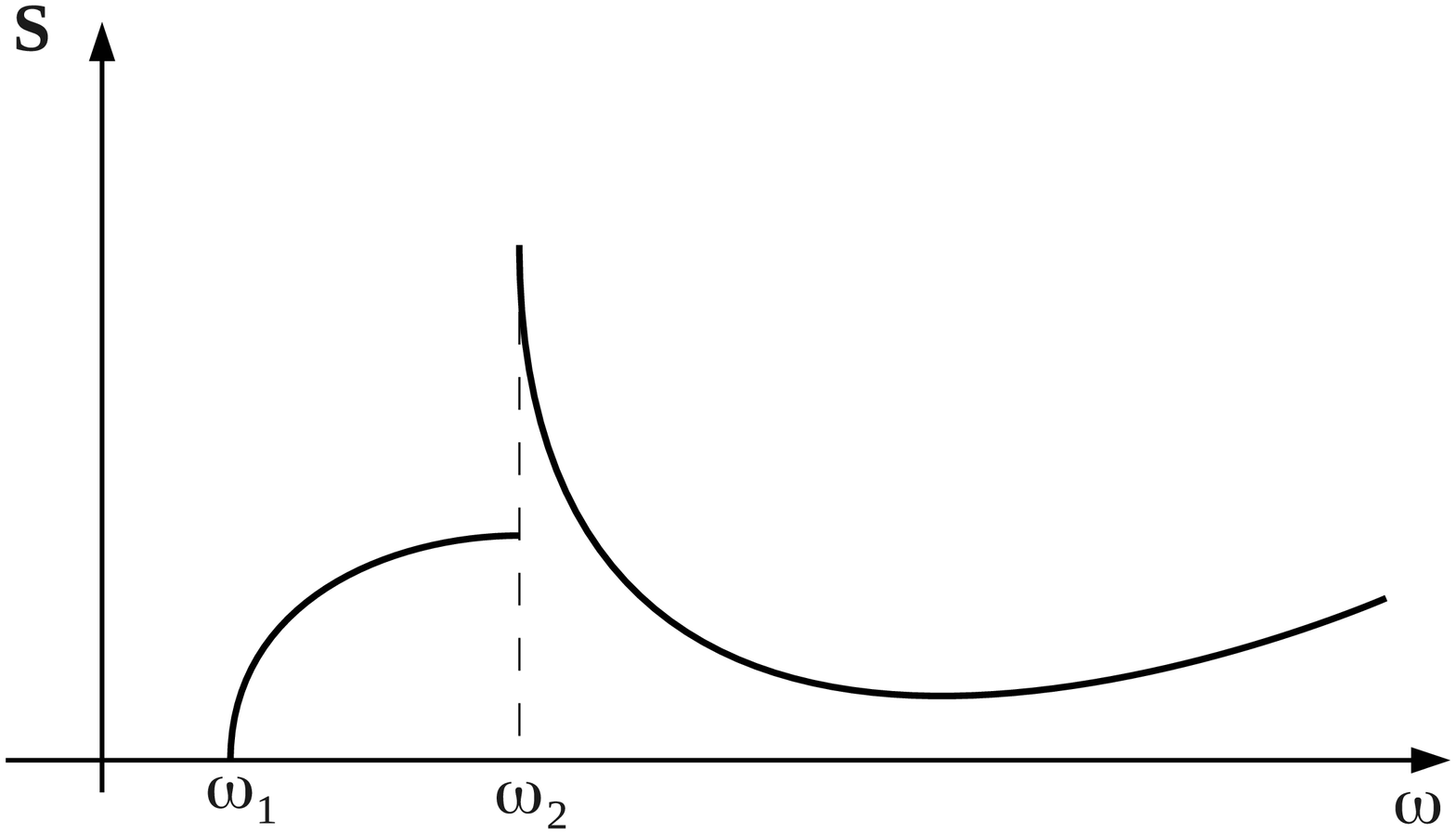}}\\
  \subfloat[$|q|\lesssim H$ for $S^{-+}$]{\label{fig:gull1}\includegraphics[scale=.3]{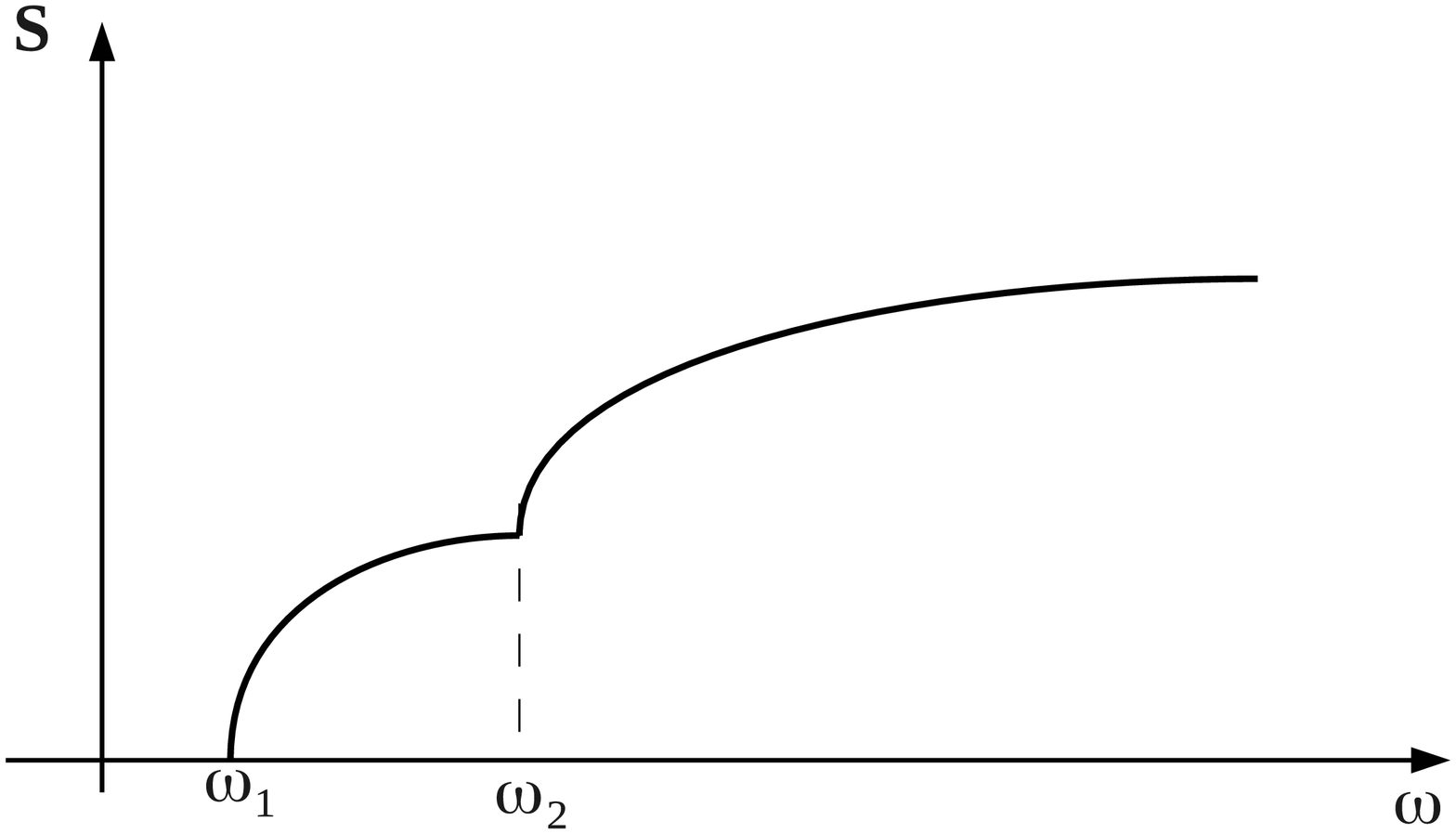}}                
  \subfloat[$H\lesssim |q|$ for $S^{-+}$]{\label{fig:tiger1}\includegraphics[scale=.3]{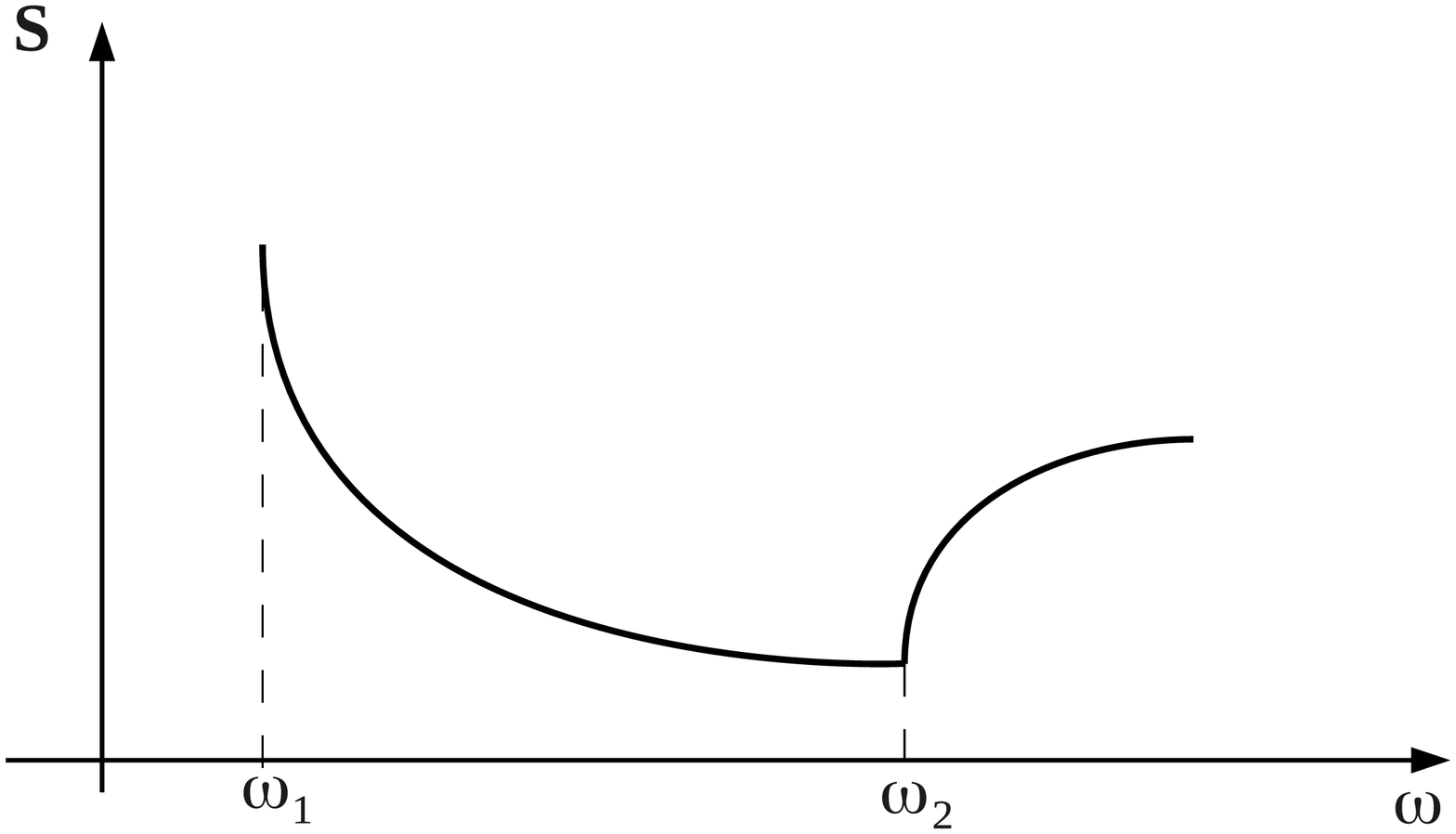}}
    \caption{Various shapes of singularities of $S^{+-}$ and $S^{-+}$ for different range of momentum $q$, with $|q|, H \ll 1$,  predicted by bosonization.}
  \label{fig:singular}
\end{figure}

In fact, as shown above $S^{+-}$ and $S^{-+}$ are a sum of two terms each with a separate threshold for all $q\neq 0$. Depending on which spectral function we look at and the value of $q$, 
these thresholds can be both diverging, both vanishing or one diverging, one vanishing. The various shapes of $S(q,\omega )$ are sketched in Fig. [\ref{fig:singular}]. In the special 
case $q=0$, there is a single term of diverging threshold type. 

For the  transverse Green's function at $q \approx \pi$ we have
\begin{equation}
<S^{+}(x,t)S^-(0)> \propto <S^{-}(x,t)S^+(0,0)> \propto  \frac{e^{i\pi x}}{(vt -x -i\epsilon)^{1/4K}(vt +x -i\epsilon)^{1/4K}}
\end{equation}
where $x=j$. 
By taking the Fourier transform we obtain the transverse spectral function near $q=\pi$:
\begin{equation}
S^{+-}(q,\omega) \ \approx \: S^{-+}(q,\omega) \: \propto \: \frac{\theta(\omega-v|q-\pi|)}{(\omega^2 - v^2(q-\pi)^2)^{1-1/4K}}
\end{equation}
So we see that the singularity exponent for staggered part is $1-1/4K$, which is different than the exponent for uniform part. 
Now there is a single  diverging threshold at $\omega = v|q-\pi |$ for either sign of $q-\pi$ for both $S^{+-}$ and $S^{-+}$.  It seems natural to assume that the diverging 
thresholds of $S^{+-}$ and $S^{-+}$ starting at $q=\pm H$ can be interpolated to the diverging thresholds terminating at $q=\pi$.  
Such an assumption goes beyond the standard bosonization approach which is restricted to $q$ near 0 and $\pi$ (and to small $H$) 
but we shall see in the next section, using X-ray edge methods, that this interpolation is correct.  However we will also find 
that the exponents of the diverging thresholds are predicted by standard bosonization: $1-2\eta$ near $|q|=H$ and $1-1/(4K)$ near $q=\pi$ are both incorrect, as are the exponents of the vanishing thresholds.  

\subsection{Effect of irrelevant band curvature operators}
\label{sec:band}
In previous sub-section we reviewed the prediction of bosonization for the singularity exponent of transverse spectral functions. In this section by including the effect of band curvature operators we show that the predictions of naive bosonization for singularity exponents are not reliable.

In bosonization approach we treat the interactions exactly but we linearize the dispersion around Fermi points and neglect the effects of higher order band curvature terms. By power
counting these terms are irrelevant in low energies and renormalize to zero, but as discussed in [\onlinecite{rodrigo07}], the effect of these  operators is important
 near the singular thresholds for longitudinal
 spectral function. We will show that the same argument works for transverse spectral function. In this section
we look at the effect of these terms to lowest order for $h \neq 0$. 

As shown in [\onlinecite{rodrigo0,rodrigo07}]  by including the effect of band curvature corrections the Hamiltonian becomes
\begin{equation}
 \mathcal{H}=\mathcal{H}_{LL}+\delta \mathcal{H}
\end{equation}
 where $\mathcal{H}_{LL}$ is Luttinger Liquid Hamiltonian and $\delta \mathcal{H}$ is given by
\begin{eqnarray}
\label{pertH}
 \delta \mathcal{H} & = &\frac{\sqrt{2\pi}}{6}\int dx \{ \eta_- [(\partial_x\varphi_L)^3-(\partial_x\varphi_R)^3] \nonumber \\
&+& \eta_+ [(\partial_x\varphi_L)^2\partial_x\varphi_R-(\partial_x\varphi_R)^2\partial_x\varphi_L]\}
\end{eqnarray}
 Where to first order in $\Delta$ we have 
\begin{eqnarray*}
\eta_- & \approx & \frac{1}{m}\left(1+\frac{2\Delta}{\pi}\sin k_F\right) \\
\eta_+ & \approx & -\frac{3\Delta}{\pi m}\sin k_F
\end{eqnarray*}
Where $m=(J \cos k_F)^{-1}$ is the effective mass of Fermi excitations. For weak interaction we can neglect $\eta_+$ and only 
include the effect of 
$\eta_-$ term in  Eq. (\ref{pertH}). \\

\begin{figure}[htp]
\includegraphics[scale=.7]{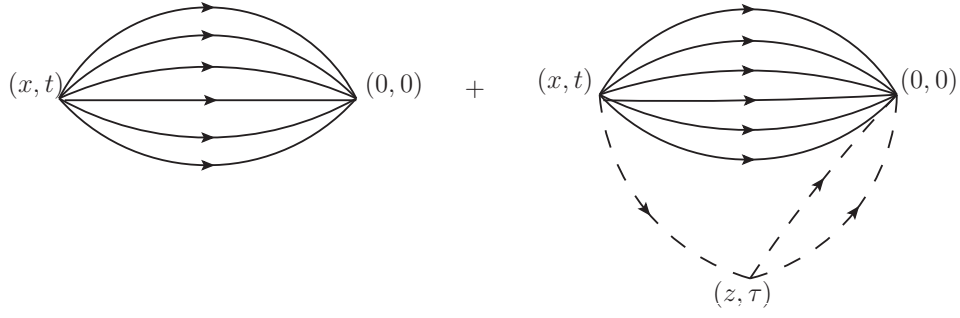}
\caption{The relevant Feynman graphs which contribute to correlation in first order of perturbation.}
\label{feynman}
\end{figure}

Now let us evaluate the transverse spectral function using perturbation theory in  $\delta \mathcal{H}$. 
By ignoring terms proportional to $\eta_+$ which mixes right and left operators, in general we are looking for the following kind of imaginary time correlation functions
\begin{eqnarray}
 G_{\nu \bar{\nu}}(x,\tau ) & = & G_R(x,\tau ) G_L(x,\tau ) \nonumber \\
G_R(x,\tau ) & = & <e^{i\sqrt{2\pi\nu}\varphi_R(x,\tau )}e^{-i\sqrt{2\pi\nu}\varphi_R(0,0)}>  \nonumber \\
G_L(x,\tau ) & = & <e^{i\sqrt{2\pi\bar{\nu}}\varphi_L(x,\tau )}e^{-i\sqrt{2\pi\bar{\nu}}\varphi_L(0,0)}> 
\end{eqnarray}
Where $\nu$ and $\bar{\nu}$ could be written explicitly in terms of Luttinger parameter, $K$, but for the following discussion we do not need their explicit form. The first order correction from perturbation Eq. (\ref{pertH}) modifies the correlation function to 
\begin{eqnarray}
\label{perturbation}
 G_{\nu \bar{\nu}}(x,\tau ) & = & G^{(0)}_R(x,\tau )  G^{(0)}_L(x,\tau )\nonumber \\
 & - & \frac{i\eta_- \sqrt{2\pi}}{6} \int dz d\tau '<e^{i\sqrt{2\pi\nu}\varphi_R(x,\tau )}(\partial_z 
\varphi_R(z,\tau '))^3e^{-i\sqrt{2\pi\nu}\varphi_R(0,0)}>_0 G^{(0)}_L(x,\tau )  \nonumber \\
   & + & \frac{i\eta_- \sqrt{2\pi}}{6} \int dz d\tau '<e^{i\sqrt{2\pi\bar{\nu}}\varphi_L(x,\tau )}(\partial_z 
\varphi_L(z,\tau '))^3e^{-i\sqrt{2\pi\bar{\nu}}\varphi_L(0,0)}>_0 G^{(0)}_R(x,\tau ) \nonumber \\
\end{eqnarray}
To evaluate the correlation function Eq. (\ref{perturbation}), we focus on corrections to $G_R(x,\tau )$; calculations for 
$G_L(x,\tau )$ are exactly the same. \\
So we have 
\begin{eqnarray}
 G_R(x,\tau ) & = & G^{(0)}_R(x,\tau ) \nonumber \\
&  - & \frac{i\eta_- \sqrt{2\pi}}{6} \int dz d\tau '<e^{i\sqrt{2\pi\nu}\varphi_R(x,\tau )}(\partial_z \varphi_R(z,\tau '))^3e^{-i\sqrt{2\pi\nu}\varphi_R(0,0)}>_0 G^{(0)}_L(x,\tau )
\end{eqnarray}
In diagrammatic way the non-zero contributions are depicted in Fig. [\ref{feynman}] . \\

So the non-zero correction is given by 
\begin{eqnarray}
\delta G_R & \propto & \int dzd\tau ' <e^{i\sqrt{2\pi\nu}\varphi_R(x,\tau )}(\partial_z \varphi_R(z,\tau '))^3e^{-i\sqrt{2\pi\nu}\varphi_R(0,0)}> \nonumber \\
& = & \int dzd\tau '\int \sum_{n,m} \frac{(i\sqrt{2\pi \nu})^n (-i\sqrt{2\pi \nu})^m}{n! m!}<\varphi_R(x,\tau )^n(\partial_z\varphi_R(z,\tau '))^3 \varphi_R(0,0)^m > \nonumber \\
& = & \int dzd\tau ' \sum_{m,n}3 n \: m(m-1)\frac{(i\sqrt{2\pi \nu})^n (-i\sqrt{2\pi \nu})^m}{n! m!} <\varphi_R(x,\tau )^{n-1}
\varphi_R(0,0)^{m-2}> \nonumber \\
&\times & <\varphi_R(x,\tau )\partial_z\varphi_R(z,\tau ')><\partial_z\varphi_R(z,\tau ')\varphi_R(0,0)>^2
\end{eqnarray}
In going from second line to third line, we have used Wick's theorem, and the factor $3n\: m (m-1)$ comes from the all possible number of contraction of fields. Thus we have
\begin{eqnarray}
\label{delta}
\delta G_R & \propto & -3i \sqrt{(2\pi \nu)^3}G^{(0)}_R(x,\tau )\int dzd\tau '   <\varphi_R(x,\tau )
\partial_z\varphi_R(z,\tau ')><\partial_z\varphi_R(z,\tau ')\varphi_R(0,0)>^2
\end{eqnarray}
Now by using the fact that 
\begin{equation}
 <\partial_x\varphi_{R,L}(x,\tau )\varphi_{R,L}(0,0)> = \frac{1}{2\pi}\frac{1}{v\tau \mp i x}
\end{equation}
we can write Eq. (\ref{delta}) as 
\begin{eqnarray}
\delta G_R & \propto & 3i \sqrt{(2\pi \nu)^3}G^{(0)}_R(x,\tau )\int dz\: d\tau  '
\frac{1}{v(\tau '-\tau )-i(z-x)}\frac{1}{(v\tau '-iz)^2} \nonumber \\
& = &  -12 \pi i \sqrt{(2\pi \nu)^3}\frac{1}{(v\tau -ix)^{\nu}} \int d\tau '\frac{\hbox{sgn}(\tau ')-\hbox{sgn}(\tau '-\tau )}
{(v\tau -ix)^2} \nonumber \\
& = &  -12 \pi i \sqrt{(2\pi \nu)^3} \frac{t}{(v\tau -ix)^{2+\nu}}
\end{eqnarray}
By replacing $v\tau  \rightarrow ((v\tau -ix)+(v\tau +ix))/2$ we can write $\delta G_R$ as
\begin{eqnarray}
 \delta G_R & \propto -\frac{6 \pi i \sqrt{(2\pi \nu)^3}}{v} \left(\frac{1}{(v\tau -ix)^{1+\nu}}+\frac{v\tau +ix}{(vt-ix)^{2+\nu}}\right)
\end{eqnarray}
 With exactly the same calculations we will get results for left moving fields as
\begin{eqnarray}
 \delta G_L & \propto -\frac{6 \pi i \sqrt{(2\pi \bar{\nu})^3}}{v} \left(\frac{1}{(v\tau +ix)^{1+\bar{\nu}}}+\frac{v\tau -ix}
{(v\tau +ix)^{2+\bar{\nu}}}\right)
\end{eqnarray}
Now by plugging all these results into Eq. (\ref{perturbation}), the final form of the correlation function to first order is given by
\begin{eqnarray}
\label{corr}
 G_{\nu \bar{\nu}}(x,\tau ) & = & \frac{1}{(v\tau -ix)^{\nu}(v\tau +ix)^{\bar{\nu}}} \{1-\frac{4\pi^2 \eta_- \sqrt{\nu^3}}{v}\left(\frac{1}{v\tau -ix}+\frac{v\tau +ix}{(v\tau -ix)^{2}}\right) \nonumber \\
& + & \frac{4\pi^2 \eta_- \sqrt{\bar{\nu}^3}}{v}\left(\frac{1}{v\tau +ix}+\frac{v\tau -ix}{(v\tau +ix)^{2}}\right)\}
\end{eqnarray}
Now by taking the Fourier transform of Eq. (\ref{corr}) and continuing to real frequencies,  we have
\begin{eqnarray}
 \label{fcorr}
 G_{\nu \bar{\nu}}(\omega,q) & = & (\omega-vq)^{\nu-1}(\omega+ vq)^{\bar{\nu}-1}\{1-\frac{4\pi^2 \eta_- \sqrt{\nu^3}}{v}(\omega-vq)\left(1+\frac{\omega-vq}{\omega+vq}\right) \nonumber \\
& + & \frac{4\pi^2 \eta_- \sqrt{\bar{\nu}^3}}{v}(\omega+vq)\left(1+\frac{\omega+vq}{\omega-vq}\right)\}
\end{eqnarray}
Where $\eta_- \approx 1/m$. It is easily seen from the last term of Eq. (\ref{fcorr}) that, as $\omega$ approaches $ vq$, the perturbative corrections blow up like $q^2/m(\omega-v q)$. This is exactly the reason that
bosonization fails near the threshold. We also see that for $|\omega-vq|\gg q^2/2m$,
 perturbative corrections  become small and we get the naive bosonization results, so we showed that irrelevant operators potentially will change the singularity
 exponent of correlation functions near the threshold 
but their effect is negligible away from the threshold so they would not change the qualitative shape of correlation function found by bosonization. In the following section by using the X-Ray edge method we find the singularity exponents of transverse spectral functions.

\section{X-Ray Edge method}
\label{sec:xray}
In sub-section \ref{sec:bosonize}, by use of standard bosonization we found that the transverse spectral functions $S^{+-}(\omega,q)$, for $q$ near $0$ 
 has a diverging singularity with exponent $1-2\eta$. In sub-section \ref{sec:band} we argued that band curvature operators would change the result of bosonization for singularity exponents.
In this section we will  explore this question by extending bosonization using X-ray edge methods introduced in [\onlinecite{glazman09,glazman06,rodrigo08,rodrigo09} ].
 We find that by use of these methods, transverse spectral functions 
 have different critical  exponents at singular energies  than predicted by standard bosonization.  We also point out the existence of large numbers of sub-leading singularities with vanishing intensities, similar 
 to the ones found by standard bosonization.  We use the notation and results of [\onlinecite{rodrigo09}] and we skip 
the details of derivations; interested readers should see [\onlinecite{rodrigo09,rodrigo08}] for most detailed calculations. 
In this approach we try to evaluate transverse spectral functions for fixed momentum $q$ by including the effect of a single high-energy particle or hole excitation. To do this we need to find relevant momenta of this excitation which contributes to the spectral functions at momentum $q$. In this section we find the effective Hamiltonian for these excitations, and also those relevant momenta.
Suppose that the momentum of this particle or hole excitation is $k$.  (We will eventually use the notation $k_p$ for a particle and $k_h$ for a hole.) In X-Ray edge method we are interested in the high energy excitations near this momentum $k$ and also low-energy excitations around the fermi points. Thus we can write the fermion operator in the following form 
\be c_j\approx \psi_R e^{ik_Fj}+\psi_L e^{-ik_Fj}+d\: e^{ikj}. \label{psi} \ee
Where $\psi_R, \psi_L$ and $d$ vary slowly on lattice scale. Then by linearizing the dispersion around the fermi points and bosonizing low energy fermions, and also linearizing the dispersion around high-energy particle 
or hole excitation, we have 
\be
\label{linear}
 \mathcal{H} = d^{\dagger} \left(\epsilon - i u\partial_x  \right) d \nonumber \\
 +  \frac{v}{2}\left[(\partial_x \varphi_L)^2+(\partial_x \varphi_R)^2\right] \nonumber \\
+  \frac{1}{\sqrt{2\pi K}} \left(\kappa_L \partial_x \varphi_L -\kappa_R \partial_x \varphi_R\right)d^{\dagger} d
\ee
Here the chiral fields $\varphi_L$ and $\varphi_R$ are the transformed ones defined by:
\begin{eqnarray}
\phi & = & \frac{\varphi_{L}-\varphi_{R}}{\sqrt{2}} \nonumber \\
\theta& = & \frac{\varphi_{L}+\varphi_{R}}{\sqrt{2}} .\nonumber
\end{eqnarray}
 This Hamiltonian is described in [ \onlinecite{neto,balents,tfk}] for Luttinger liquid coupled to an impurity. The parameters of above hamiltonian are as follows; $\epsilon$ is the 
energy of the high energy particle or hole and, for $\Delta =0$, it is given by $\epsilon =-2J (\cos k-\cos k_F)$. 
$u$ is the velocity of the heavy particle or hole and equals $ J \sin k$ at $\Delta =0$.  $v$ and $K$ are  the boson  velocity and Luttinger 
parameter,  respectively.  $v$ may be regarded as the Fermi velocity of the interacting fermion model. It is the only velocity appearing in the standard bosonization 
approach and plays the role of the ``velocity of light'' in the effective Lorentz invariant field theory. The  velocity parameter $u$, describing the 
high energy particle or hole is an important new parameter in the X-ray edge approach. 
Finally $\kappa_{R,L}$ are the couplings between high energy fermion
 and bosonic fields and to first order in $\Delta $ are given by $\kappa_{R,L}= 2\Delta [1-\cos(k_F \mp k)]$.  These coupling could be evaluated by 
Bethe ansatz calculations\cite{korepin0} for any $\Delta$,  $H$ and $k$. They 
 are important for finding the singularity exponents. Note that, in general, 
$u$, $\epsilon$ and $\kappa_{R,L}$ all depend on $k$ as well as $\Delta$ and $h$. 

The Hamiltonian (\ref{linear}) looks complicated as it contains interactions between fermions and bosons. We can eliminate the interacting part of (\ref{linear}) by doing 
a unitary transformation given by 
\begin{equation}
 \label{unitary}
U = \exp \left[ -i \int \frac{dx}{\sqrt{2\pi K }}\left(\gamma_R\varphi_R + \gamma_L\varphi_L\right)d^{\dagger} d\right]
\end{equation}
Effect of unitary transformation on bosonic and fermionic fields is as follows
\begin{eqnarray}
\label{transform}
 \partial_x \varphi_{R,L} & = & \partial_x \bar{\varphi}_{R,L}\pm \frac{\gamma_{R,L}}{\sqrt{2\pi K}}\bar{d}^{\dagger}(x) \bar{d}(x) \nonumber \\
   d& = & \bar{d}\: \exp \left[\frac{-i}{\sqrt{2\pi K}}(\gamma_R \bar{\varphi}_R+\gamma_L\bar{\varphi}_L)\right] \nonumber \\
 \varphi_{R,L} & = & \bar{\varphi}_{R,L} \pm  \frac{\gamma_{R,L}}{4 \sqrt{2\pi K}} \tilde{N}(x) \nonumber 
\end{eqnarray}
where $ \tilde{N}(x)$ is defined by:
\be   \tilde{N}(x)  =  \int_{-\infty}^{\infty}\hbox{sgn}(x-y)\bar{d}^{\dagger}(y) \bar{d}(y) dy.\ee
It is easy to see that unitary transformation leaves $d^{\dagger}(x)d(x)$ invariant. With $\gamma_{R,L}$  given by
\begin{equation}
 \label{gamma}
\gamma_{R,L} = \frac{\kappa_{R,L}}{v\mp u}
\end{equation}
we can decouple fermionic fields from the bosonic ones. Having done the unitary transformation the Hamiltonian will look like
\begin{equation}
 \mathcal{H} = \frac{v}{2}\left[(\partial_x \bar{\varphi}_L)^2+(\partial_x  \bar{\varphi}_R)^2\right] +
\bar{d}^{\dagger} \left(\epsilon - i u \partial_x  \right) \bar{d} + \cdots
\end{equation}
Where $\cdots$ means higher dimension irrelevant interactions that will be produced by doing unitary transformation and which we ignore. 

We now consider the transverse Green's function:
\[S^{-+}=<S^-_j(t)S^+_0(0)>.\]
By doing Jordan-Wigner transformation,  we have
\[S^{-+}_j(t)=e^{i\pi j}<c_j(t) \cos [\pi N_j(t)]\cos [\pi N_0(0)]c_0^{\dagger}(0)>\]
where $c_j(t)$ is approximated as in Eq.  (\ref{psi}) and 
\[N_j(t)=\sum_{l<j}c^{\dagger}_l(t)c_l(t)\]
To obtain the transverse spectral function, $S^{-+}(q,\omega )$ at a wave-vector $q$ far from the low energy regions, $\pm H$, $\pi$, 
 the term that we are interested in is
\be \label{first+-} S^{-+}_j(t)=e^{i\pi j}e^{ik_p j}<d(j,t)\cos [\pi N_j(t)]\cos [\pi N_0(0)]d^{\dagger}(0,0)>.\ee
We see that $d$ must be chosen to be a particle operator and we have consequently labelled its momentum $k_p$. 
Note that we have written the Jordan-Wigner string operator in manifestly Hermitian $\cos$ form, as in Sec. IIIA. 
Now  we  decompose $N_j(t)$, into  c-number, non-oscillatory and oscillatory parts
\begin{eqnarray*}
\label{N}
N_j(t) & =&\frac{k_F}{\pi}j + \tilde{n}(j,t)+m(j,t)\nonumber \\
  \tilde{n}(x,t) & = &\int_{\infty}^{x-\epsilon} dy :\psi_R^{\dagger}(y,t)\psi_R(y,t):+:\psi_L^{\dagger}(y,t)\psi_L(y,t):+d^{\dagger}(y,t)d(y,t) \nonumber \\
m(x,t) &= &\int_{\infty}^{x-\epsilon} \left[ \psi_R^{\dagger}(y,t)d(y,t) e^{i(k_p-k_F)y}+\psi_L^{\dagger}(y,t)d(y,t)e^{i(k_F+k_p)y}+\psi^\dagger_L(y,t)\psi_R(y,t)e^{2ik_Fy}+h.c.\right]  \nonumber \\
  & \propto & \frac{1}{i(k_p-k_F)}  \psi_R^{\dagger}(x',t)d(x',t) e^{i(k_p-k_F)x} + \frac{1}{i(k_F+k_p)}\psi_L^{\dagger}(x',t)d(x',t)e^{i(k_F+k_p)x}+
    \frac{1}{2ik_F}\psi^\dagger_L(x',t)\psi_R(x',t)e^{2ik_Fx}+h.c.
\end{eqnarray*}
where $\epsilon \to 0^+$, $x'\equiv x-\epsilon$ and in the third line we used the fact that both $\psi$ and $d$ are slowly varying fields and most of the contribution of
the integral  comes from limiting point $x-\epsilon$.  At this point we will set the rapidly oscillating term, $m(x,t)$, to zero. This will give the dominant 
divergent singularity in the transverse spectral function.  By Taylor expanding in powers of $m(x,t)$ we obtain various vanishing singularities 
as well as unimportant renormalizations of the amplitude of the divergent singularity, as we discuss in Sec IV.  We may then decompose 
$\tilde n(x,t)$ into its commuting high energy and Fermi surface part.
\bea \tilde n(x,t)&\equiv& n(x,t)+n_d(x,t)\nonumber \\
n(x,t)&\equiv&\int_{\infty}^{x-\epsilon}:\psi_R^{\dagger}(y,t)\psi_R(y,t):+:\psi_L^{\dagger}(y,t)\psi_L(y,t):\nonumber \\
n_d(x,t)&\equiv&\int_{\infty}^{x-\epsilon} d^{\dagger}(y,t)d(y,t) dy.\nonumber \eea
Because all the $d$ operators in $\tilde{n}(x,t)$ are at points $y<x$, we have 
\[[\tilde{n}(x,t),d(x,t)]=0\]
and thus we may drop the $n_d$ terms leaving:
\[S^{-+}_j(t)\propto  e^{i(k_p +\pi)j}<\cos [k_Fj+\pi  n(j,t)]d(x,t)d^{\dagger}(0,0)\cos [\pi n(0,0)]>\]
Following Eq.  (\ref{contstring}) this becomes:
\[S^{-+}_j(t)\propto  e^{i(k_p +\pi)j}<\cos [k_Fj+\sqrt{\pi K}\phi (j,t)] d(j,t)d^{\dagger}(0,0)\cos [\sqrt{\pi K}\phi (0,0)]>.\]
Note that we have treated the Jordan-Wigner string operator in precisely the same approximation as in the standard bosonization approach. 
We now make the  unitary transformation of Eq.  (\ref{transform}) so that the fermion and bosons are decoupled. Noting that $\tilde N(x,t)$ annihilates 
the vacuum this leaves:
\begin{eqnarray}
\label{decouple}
 S^{-+} & \propto  &<\bar{d}(x,t)\cos [k_F x+\sqrt{\pi K} \bar{\phi}(x,t)]
e^{\frac{-i}{\sqrt{2 \pi K}}[\gamma_R \bar{\varphi}_R(x,t)+\gamma_L\bar{ \varphi_L}(x,t)]}\nonumber \\
&\:& e^{\frac{i}{\sqrt{2 \pi K}}[\gamma_R \bar{\varphi}_R(0,0)+\gamma_L\bar{ \varphi_L}(0,0)]}
\cos [\sqrt{\pi K} \bar{\phi}(0,0)]\bar{d}^{\dagger}(0,0)>e^{i(k_p+\pi)x} \nonumber 
\end{eqnarray}

Separating the fermionic and bosonic factors, this becomes:
\be S^{-+}_j(t)=S^{(0)-+}(x,t)<\bar{d}(x,t)\bar{d}(0)^{\dagger}>\label{S-+d}\ee
where $x=j$, 
\be S^{(0)-+}(x,t)=e^{i(\pi+k_p-k_F)x} I^-(x,t)+e^{i(\pi+k_p +k_F )x} I^+(x,t)\label{mom}\ee
and 
\begin{eqnarray}
 I^+(x,t) & = & \left<e^{-i\sqrt{2\pi \nu^+_R}\bar{\varphi}_R(x)+i\sqrt{2\pi \nu^+_L}\bar{\varphi}_L(x)}e^{i\sqrt{2\pi \nu^+_R}\bar{\varphi}_R(0)-i\sqrt{2\pi \nu^+_L}\bar{\varphi}_L(0)} \right> \nonumber \\
& = & \left(\frac{\epsilon}{\epsilon+ivt-ix}\right)^{\nu^+_R}\left(\frac{\epsilon}{\epsilon+ivt+ix}\right)^{\nu^+_L} \nonumber \\
I^-(x,t) & = & \left<e^{i\sqrt{2\pi \nu^-_R}\bar{\varphi}_R(x)-i\sqrt{2\pi \nu^-_L}\bar{\varphi}_L(x)}e^{-i\sqrt{2\pi \nu^-_R}\bar{\varphi}_R(0)+i\sqrt{2\pi \nu^-_L}\bar{\varphi}_L(0)}  \right>  \nonumber \\
& = & \left(\frac{\epsilon}{\epsilon+ivt-ix}\right)^{\nu^-_R}\left(\frac{\epsilon}{\epsilon+ivt+ix}\right)^{\nu^-_L}\label{I}
\end{eqnarray}
Where $\epsilon$ is of order of the lattice spacing  and  $\nu^{\pm}_{R,L}$ are defined as follow
\begin{eqnarray}
\label{zeroexp}
\nu^{\pm}_{R} & = & \frac{1}{4}\left(\frac{\gamma_{R}}{\pi \sqrt{K}}\pm \sqrt{K}\right)^2 \nonumber \\
\nu^{\pm}_{L} & = &\frac{1}{4}\left(\frac{\gamma_{L}}{\pi \sqrt{K}}\mp \sqrt{K}\right)^2.
\end{eqnarray}
At zero magnetic field we have $\gamma_{R,L}/\pi=1-K$,\cite{rodrigo08} independent of momentum, so the results will simplify to
\bea \nu^{\pm}_{R}&=&\frac{1}{4K}\left(1-K \pm K\right)^2\nonumber \\
\nu^{\pm}_{L}&=&\frac{1}{4K}\left(1-K \mp K\right)^2.\label{h0exp}\eea
Having done the unitary transformation, the $\bar d$ fields act as free particle so we have
\begin{equation}
 <\bar d(x,t)\bar d^{\dagger}(0)> \approx e^{-i \epsilon(k_p) t} \int_{-\Lambda}^{\Lambda} \frac{dk}{2\pi} e^{ik(x-ut)} \approx e^{-i \epsilon(k_p) t} \delta(x-ut).\label{d}
\end{equation}

We can now turn to the question of how the momentum of the high energy particle, $k_p$, should be chosen to study $S^{-+}(q,\omega )$ near 
the threshold for arbitrary $q$.  From Eq.  (\ref{mom}) we see that a high energy particle of momentum $k_p$ gives terms in the transverse 
Green's function oscillating at wave-vectors $\pi +k_p-k_F$ and $\pi + k_p+k_F$.   Thus we see that there may actually be two choices for $k_p$ which 
will give a contribution to the transverse Green's function oscillating at a specified wave-vector $q$:
\be k_p^{\pm}\equiv q+\pi \mp k_F.\ee
In general, both must be considered in calculating the singularity behaviour of $G^{-+}(q,\omega )$. 
However, the  $k_p$'s are restricted by the requirement that they are allowed particle momenta, 
\be k_F<k_p<2\pi -k_F,\  \  (\hbox{mod}\  2\pi ) .\ee
(Recall that $k_F=\pi /2+\pi m>\pi /2$.)
Thus we see that the high energy particle of momentum $k_p^+$ contributes to  $S^{-+}(q,\omega )$ for $q$ in the range  $[2k_F-\pi ,\pi ]$ and 
the particle of momentum $k_p^-$ contributes for $q$ in the range $[-\pi,\pi-2k_F]$.  For a given $q$ there is at  most
one possible
high energy particle momentum contributing to $S^{-+}(q,\omega )$. Having identified the appropriate high energy particle momentum we may 
now complete the calculation by Fourier transforming Eq.  (\ref{S-+d}) using Eqs.  (\ref{I})-(\ref{d}).  The result is:
\bea S^{-+}(q,\omega )&\propto& \int dx dt {e^{i[\omega -\epsilon (k_p^+)]t}\delta (x-ut)
\over (vt-x-i\epsilon)^{\nu_R^+}(vt+x-i\epsilon)^{\nu_L^+}}, \ \  (2k_F-\pi <q<\pi )   \nonumber \\
&\propto& \int dx dt {e^{i[\omega -\epsilon (k_p^-)]t}\delta (x-ut)
\over (vt-x-i\epsilon)^{\nu_R^-}(vt+x-i\epsilon)^{\nu_L^-}} ,\ \  (-\pi <q<\pi -2k_F) .
\eea
The $x$-integrals may be done trivially using the $\delta$-functions. 
\bea S^{-+}(q,\omega )&\propto& \int dt {e^{i[\omega -\epsilon (k_p^+)]t}
\over [(v-u)t-i\epsilon]^{\nu_R^+}[(v+u)t-i\epsilon]^{\nu_L^+}} ,\  \  (2k_F-\pi <q<\pi )  \nonumber \\
&\propto& \int dt {e^{i[\omega -\epsilon (k_p^-)]t}
\over [(v-u)t-i\epsilon]^{\nu_R^-}[(v+u)t-i\epsilon)^{\nu_L^-}}, \  \   (-\pi <q<\pi -2k_F).\label{t-int}
\eea
The $t$-integrals can now be done by contour methods. Note that, if $v>u$, they are only non-zero for $\omega >\epsilon (k_p^+)$ 
and $\omega >\epsilon (k_p^-)$ respectively, corresponding to a lower threshold. Using the free particle cosine dispersion relation $v>u$ is always satisfied 
for particle excitations. While this dispersion relation is known to be exact, apart from an overall factor, including interactions for $h=0$ it is in general modified. We might expect that $v>u$ remains true for particles, at least for small enough $h$. However, see below. Assuming this, we obtain:
\be S^{-+}(q,\omega )\propto {\theta [\omega -\omega_L(q)]\over [\omega -\omega_L(q)]^{\mu (q)}}\ee
where the singular energies are given by:
\bea \omega_L(q)&=&\epsilon (k_p^+),\  \   (2k_F-\pi <q<\pi ) \nonumber \\
&=&\epsilon (k_p^-),\  \  (-\pi <q<\pi -2k_F)   .\eea
The critical exponents are given by 
\bea \mu (q)&=& 1-\nu_R^+(k_p^+)-\nu_L^+(k_p^+),\ \   (2k_F-\pi <q<\pi )    \nonumber \\
&=& 1-\nu_R^-(k_p^-)-\nu_L^-(k_p^-),\ \   (-\pi <q<\pi -2k_F) \eea
with $\nu_{R,L}^{\pm}(k)$ given by Eqs. (\ref{zeroexp}) and (\ref{h0exp}).  Note that the same phase shift parameters $\gamma_{R/L}$ determine 
both longitudinal and transverse spectral functions; however we need to know them at the momentum of the high energy 
particle or hole which is not the same for longitudinal and transverse spectral functions, for given $q$.  In general, the phase shift parameters 
depend on momentum as well as field, becoming momentum independent at $h=0$. It follows from parity that 
$\kappa_L(k)=\kappa_R(-k)$. Since, for $q>0$, $k_p^-(-q)=-k_p^+(q)$, (mod $2\pi$) $\nu_L^-(-q)=\nu_R^+(q)$ and $\nu_R^-(-q)=\nu_L^+(q)$ 
and hence $\mu (-q)=\mu (q)$. \\

It is interesting to compare both the singular energies and exponents to those predicted by standard bosonization as $q$ approaches the zero energy 
points $q\approx \pm H=\pm (2k_F-\pi )$ and $q\approx \pi$. Near these zero energy points we may linearize the $\epsilon (k)$ giving:
$ \omega_L(q)\approx v|q\mp H|$ and $v|q-\pi |$, precisely the singular energies predicted by standard bosonization. This X-ray edge calculation 
also confirms the conjecture made in Sec. II that the diverging singular energies at the different low energy momenta are smoothly connected. 
On the other hand, the exponents appear to disagree with the standard bosonization results for all $\Delta$ and $h$, a similar observation to the one in Ref. [\onlinecite{glazman08}].  This can be seen, for example.
by considering the limit $\Delta \to 0$. In this case $\gamma^{\pm}_{L,R}\to 0$, $K\to 1$ so $\nu^{\pm}_{L,R}\to 1/4$ and $\mu \to 1/2$, for all $q$ and $h$. 
On the other hand the standard bosonization result from Eq.  (\ref{eta}) and (\ref{spec}) is $\eta \to 1/8$ and hence $\mu \to 3/4$. We can also 
compare the zero field predictions for general $\Delta$. From Eq.  (\ref{h0exp}) 
\be \mu = 2-1/(2K)-K,\ \  (h=0, \forall q).\label{muh0}\ee
 On the other hand, 
standard bosonization predicts $\mu = 1-2\eta = 2-1/(4K)-K$ for $q\approx 0$ and $\mu = 1-1/(4K)$ for $q\approx \pi$.  We expect that 
standard bosonization fails to predict critical exponents correctly for the transverse spectral function, as discussed
in sub-section IIB. Given this situation, it is useful to check the SU(2) symmetric case, $h=0$, $\Delta =1$. In this 
case X-ray edge methods predict, from Eq.  (\ref{muh0}) $\nu_R^-=\nu_L^+=0$, $\nu_R^+=\nu_L^-=1/2$, $\mu = 1/2$  for the transverse spectral function  independent of $q$.  The same
exponents were found earlier\cite{rodrigo08} for the longitudinal spectral function.\cite{AG} (In this case, they agree with 
standard bosonization near $q=\pi$ but not near $q=0$.) 

\begin{figure}[htp]
\includegraphics[scale=.5]{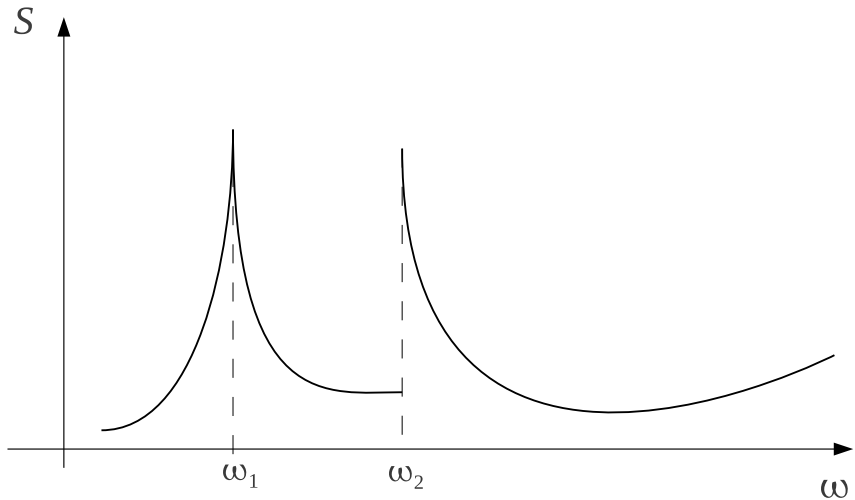}
\caption{The behaviour of $S^{+-}(\omega,q)$ correlation for $|q|<H$. There are two
different hole excitations which contribute to 
the spectral function; the lower energy hole produces a cusp-like singularity.}
\label{fig:cusp}
\end{figure}

By doing similar calculations we could find $S^{+-}$, which is different than $S^{-+}$ for $h\neq 0$. 
Following the same procedure, Eq.  (\ref{decouple}) is replaced by:
\begin{eqnarray}
\label{decouple+-}
 S^{+-} & \propto  &<\bar{d}^\dagger (x,t)\cos [k_F x+\sqrt{\pi K} \bar{\phi}(x,t)]
e^{\frac{i}{\sqrt{2 \pi K}}[\gamma_R \bar{\varphi}_R(x,t)+\gamma_L\bar{ \varphi_L}(x,t)]}\nonumber \\
&\:& e^{-\frac{i}{\sqrt{2 \pi K}}[\gamma_R \bar{\varphi}_R(0,0)+\gamma_L\bar{ \varphi_L}(0,0)]}
\cos \sqrt{\pi K} \bar{\phi}(0,0)]\bar{d}(0,0)>e^{i(-k_h+\pi)x} \nonumber 
\end{eqnarray}
where $d$ now annihilates a particle in a filled state below the Fermi energy with momentum $k_h$, i.e. creates a hole. 
This can again be factorized as:
\be S^{+-}_j(t)=S^{(0)+-}(x,t)<\bar{d}^\dagger (x,t)\bar{d}(0)>\label{S+-d}\ee
where 
\be S^{(0)+-}(x,t)=e^{i(\pi-k_h+k_F)x} I^-(x,t)+e^{i(\pi-k_h-k_F )x} I^+(x,t)\label{mom2}\ee
and $I^{\mp}(x,t)$ are the same functions defined in Eq.  (\ref{I}), except that $k_h$ must lie in a different range, $|k_h|<k_F$.  Thus defining:
\be k_h^{\mp}\equiv \pi -q\mp k_F\ee
we see that the first term in Eq.  (\ref{mom2}) is non-zero for $q$ in the range  $[\pi -2k_F,\pi ]$ while the second is non-zero for 
$q$ in the range $[-\pi, 2k_F-\pi ]$. These are wider ranges than occur in $S^{-+}$. (Recall that we assume $H\geq 0$ and hence $k_F\geq \pi /2$.)
In particular, both terms can contribute for $|q|<H$.  Again, as $q$ approaches the zero energy points, $\pm H$ and $\pi$ the singular 
energies approach those predicted by standard bosonization. Again as anticipated in sub-section IIA, the singular energies at these 
zero energy points can be smoothly connected. Another interesting feature is the shape of the singularity.  For $S^{-+}$ the 
singularity was one-sided, vanishing for $\omega <\omega_L$. This was a consequence of the fact that the velocity of the high 
energy particle always obeys $u<v$ assuming this feature of the non-interacting dispersion relation is unchanged by interactions. 
 On the other hand for holes, again using the non-interacting dispersion relation, $u<v$ is only obeyed if $|k_h|<\pi -k_F$; there is a 
range of hole momentum near $k_F$ where the high energy hole has a higher velocity than the Fermi velocity. $|k_h^+|<\pi -k_F$ 
corresponding to $0<q<2\pi -2k_F=\pi -H$. Thus, in this region the singularity is one-sided, $\propto \theta (\omega-\omega_L)$. 
On the other hand for $-H<q<0$ and $\pi -H<q<\pi$, where $u>v$ the integral in Eq.  (\ref{t-int}) is also non-zero 
and gives the same critical exponent with a different amplitude for $\omega < \omega_L$. In this case we see that 
$\omega_L(q)$ is {\it not a lower threshold}. There is also spectral weight below this frequency. The qualitative shape 
of $S^{+-}(q,\omega )$ for $0<q<H$ is sketched in Fig. [\ref{fig:cusp}] .

We emphasize that  a singularity is one-sided for $u<v$ and two-sided for $u>v$ where $v$ is the Fermi velocity 
and $u$ is the velocity of the high energy particle or hole. In general, these velocities depend on $\Delta$ and 
also $h$, being strongly renormalized by interactions.  In [\onlinecite{rodrigo07}], Fig.  [15], it was illustrated that for $\Delta =1$ 
and small non-zero field one of the cubic term in the bosonized Hamiltonian, due to band curvature effects, has a coupling 
constant $\eta_-<0$. This may indicate a reversal of the sign of the effective mass due to (strong) interaction effects, implying 
a reversal of the sign of $u-v$ as the energy of the high energy particle or hole approaches the Fermi energy, corresponding 
to $|q|\to |2k_F-\pi |$ (or $q\to \pi$). Thus in this parameter range, the one-sided and two-sided nature of the 3 singularities 
in $S^{+-}$ and $S^{-+}$, discussed above, would be reversed.

\begin{figure}
  \centering
 \subfloat[$S^{+-}$]{\label{fig:gg}\includegraphics[scale=.3]{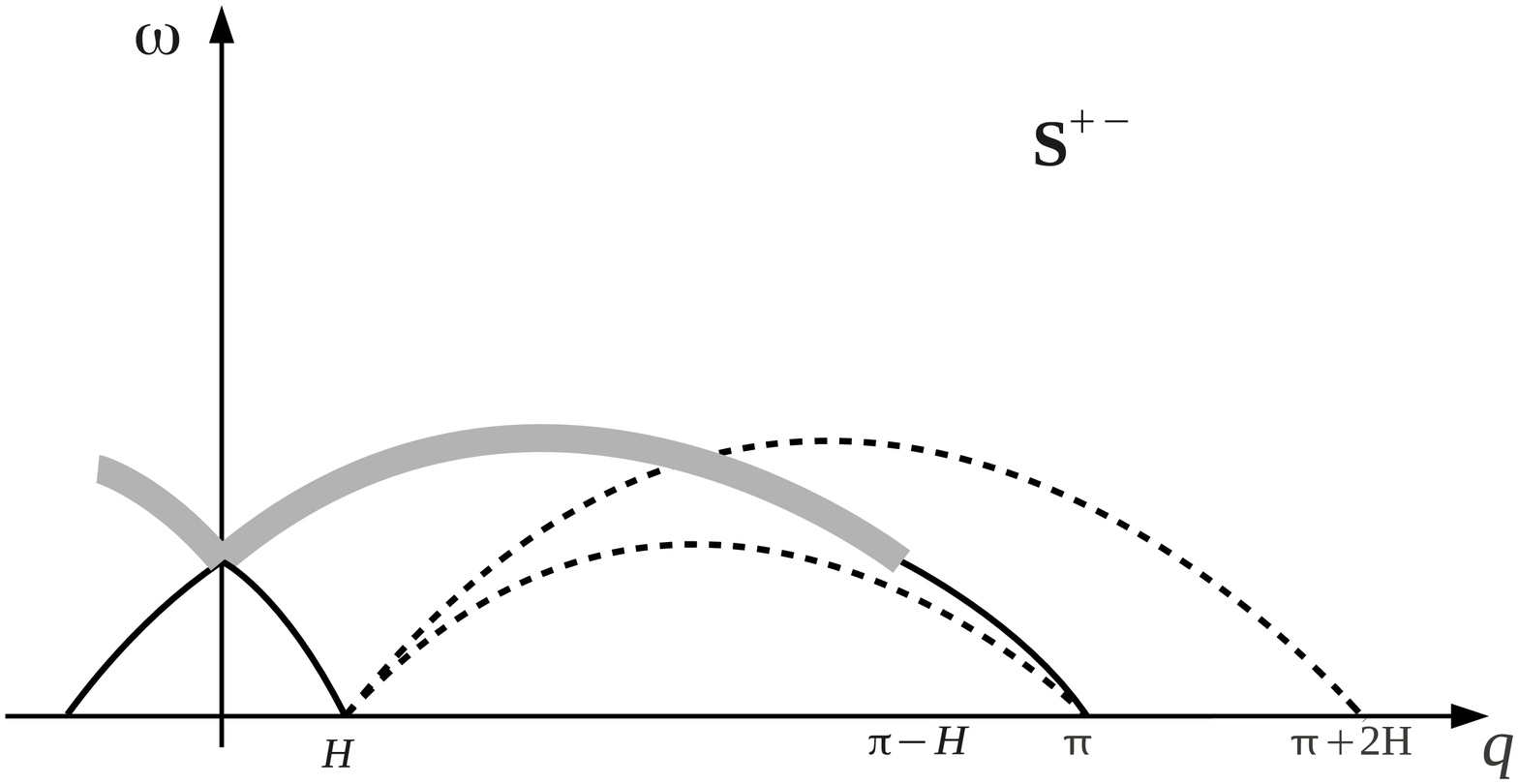}}                
 \subfloat[$S^{-+}$]{\label{fig:gg11}\includegraphics[scale=.3]{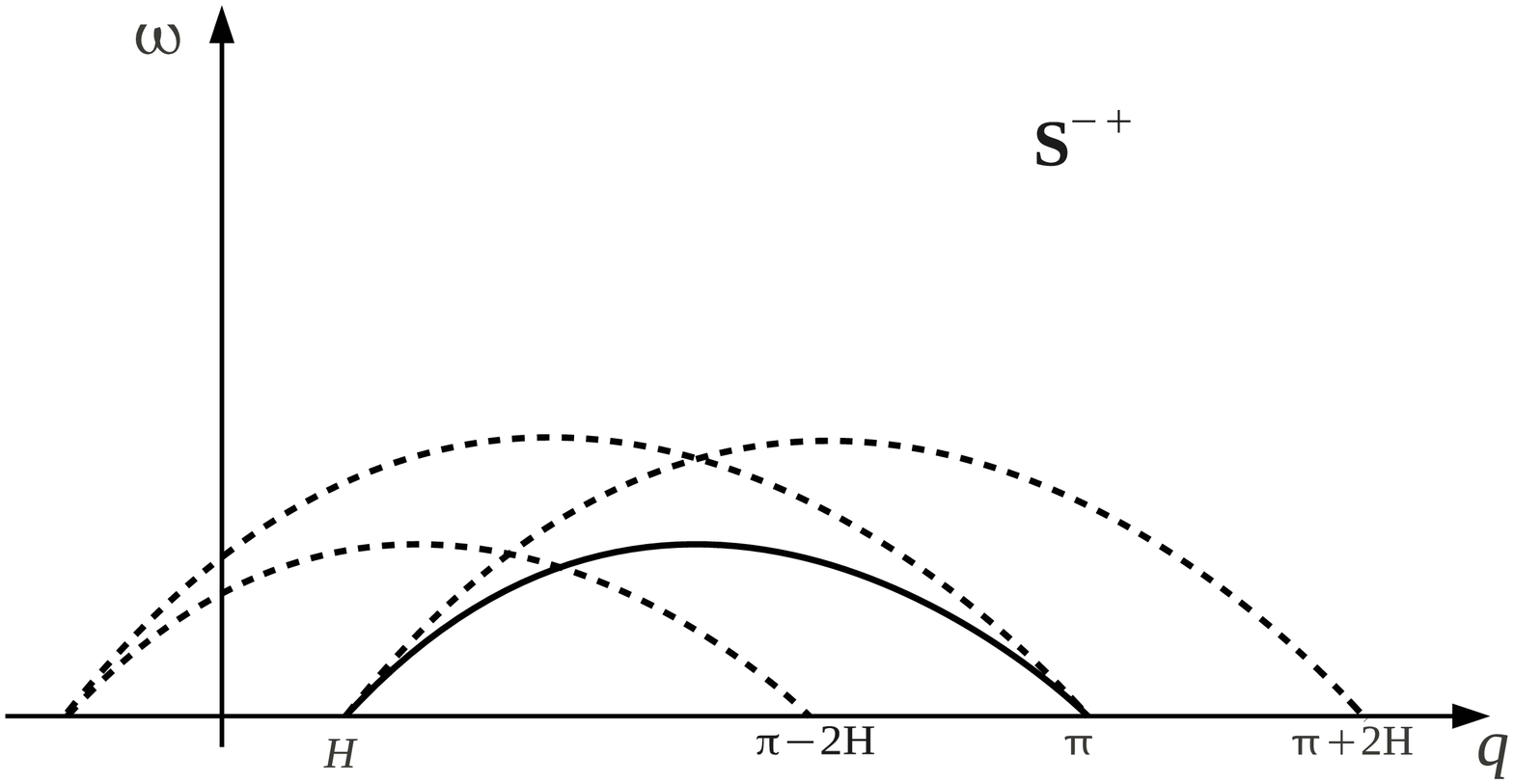}}
 \caption{Singular points of zero temperature transverse  spectral functions of the xxz
model, predicted by X-ray edge method. The solid lines indicate diverging
singularities and dashed line the vanishing singularities. The grey thick line indicates the range of momentum that interactions should broaden the diverging singularity. In Fig. [\ref{fig:gg}] and [\ref{fig:gg11}]   the lower dashed line is given by particle excitations and upper one is given by hole excitation.}
\label{fig:sing2}
\end{figure}

Given that the singular energy, $\omega_L(q)$ is sitting 
inside a region of non-zero spectral weight, for certain ranges of $q$,  we might
ask whether it is reasonable to 
expect singular behaviour at this energy or whether the infinite peak might be
broadened and made finite due to some 
sort of decay process for this high energy excitation. This important question also
arises for the longitudinal 
structure function and for the fermion spectral function.  It has been
suggested\cite{khodas,rodrigo09} that integrability might 
prevent this excitation from decaying, for some range of momentum, even though it is kinematically allowed,
leaving the singularity intact. 
This is true because three-body scattering processes are required for it to decay
and these are expected not to 
occur in this integrable model. 
This seemed to be consistent with Density Matrix Renormalization Group results for
the fermionic spectral function.\cite{rodrigo09}
In fact, we should also consider processes in which the heavy particle or hole
decays by producing 3 other high energy quasi-particles, a 2-body 
process which is  expected to be present even in this integrable model.
 Using the $-\cos k$ dispersion relation, this is kinematically allowed for high
energy holes\cite{ph} in the region 
$u<v$, corresponding to $0<|q|<\pi -H$ (the upper branch for $0<|q|<H$) but not
allowed for high energy particles for any $q$.\cite{rodrigo09}
Putting these observations together, we expect the sharp one-sided singularity of
$S^{-+}(q,\omega )$ to be present  for $H<|q|<\pi$ 
 and the  2-sided singularity of $S^{+-}(q,\omega )$ for $0<|q|<H$ and $\pi
-H<|q|<\pi$ to be present. The one-sided
singularity of $S^{+-}$ for $H<|q|<\pi -H$ 
should be broadened by higher order interaction effects\cite{rodrigo09} not 
taken into account in this treatment.  See Fig. [\ref{fig:sing2}]. Again we emphasize that the precise region of $q$ over which
singularities are broadened depends 
on the dispersion relation, which is modified by interactions; here we have just
stated it using the $-\cos k$ dispersion 
relation, valid at small $\Delta$.

So far, we have set the rapidly oscillating operator $m(x,t)$, defined in Eq. 
(\ref{N}) to zero. The effects of including 
it are discussed in the following section. It basically leads to additional terms in
the transverse spectral function which have 
singularities at different energies, including lower ones.  However, these
singularities are of vanishing type, with exponent  $\mu <0$, dashed lines of Fig. [\ref{fig:sing2}]. 
The relatively simple approach we have taken here is just sufficient to give the
diverging singular terms. 

The situation is considerably simpler at zero field, $h=0$. In this case, the free
dispersion relation is known to be exact, apart from 
an overall change of amplitude, $2t\to v$.  Thus the condition $u<v$ is always
satisfied so $S^{-+}=S^{+-}$ has only one single-sided 
singularity at $\omega_L=v\sin q$ with v given in Eq.  (\ref{vK}) and critical
exponent given by Eq.  (\ref{muh0}). 
In this case, no decay processes are kinematically 
allowed and  no additional singularities occur, since the single hole or particle
has the lowest possible energy 
for given wave-vector.

\section{Sub-dominant singularities}
In Sec. II and III we ignored rapidly oscillating terms, $m(x)$ of Eq.  (\ref{first+-}) in the Jordan-Wigner string operator in calculating the transverse structure function and also the effects of Umkalpp term which oscillates as $e^{i4k_F x}$.  We consider the effect of 
including these terms here. Let us begin with the term:
\be m(x)= \frac{1}{2ik_F}\psi^\dagger_L(x',t)\psi_R(x',t)e^{2ik_Fx}+h.c.\label{mLR}\ee
Actually, this term represents a correction to standard bosonization, even without using X-ray edge methods, so we consider its effects there. To make 
things as simple as possible we also consider zero field, $k_F=\pi /2$.  Then, after bosonizing $m$,  the standard bosonized expression for $S^-$ in Eq.  (\ref{S-low}) is modified to:
\be S^-_j\propto e^{-i\sqrt{\pi /K}\theta (x)} \left[C(-1)^j+C^-\cos (\sqrt{4\pi K}\phi (x))\right] \exp [(-1)^jiC'\sin (\sqrt{4\pi K}\phi (x))] \label{S-corr}.\ee
for a non-universal constant $C'$.  We now Taylor expand the exponential and use double angle formulas. We see that the  staggered and uniform parts 
of $S^-_j$ have a series in increasingly irrelevant operators:
\bea S^-_s&=&e^{-i\sqrt{\pi /K}\theta (x)}\sum_{n\in Z}a_{2n}e^{i2n\sqrt{4\pi K}\phi}\nonumber \\
S^-_u&=&e^{-i\sqrt{\pi /K}\theta (x)}\sum_{n\in Z}a_{2n+1}e^{i(2n+1)\sqrt{4\pi K}\phi}.\label{series}
\eea
The effect of including the $e^{im}$ factor is simply to renormalize the coefficient of the leading operator in $S^-_u$ and $S^-_s$ together 
with producing the irrelevant corrections.  Now consider non-zero field.  Eq.  (\ref{S-corr}) gets replaced by:
\be S^-_j\propto e^{-i\sqrt{\pi /K}\theta (x)} \left[C(-1)^j-C^-\cos ((2k_F-\pi )j+\sqrt{4\pi K}\phi (x))\right] \exp [iC'e^{2ik_Fj+i\sqrt{4\pi K}\phi (x)}+h.c.] \label{S-corr2}.\ee
We again get a series of irrelevant operators but now all at different wave-vectors:
\be S^-_j=e^{-i\sqrt{\pi /K}\theta (x)}\sum_{n\in Z}a_{n}e^{i(2nk_F+\pi )j+in\sqrt{4\pi K}\phi}.\label{bosexp}\ee
It is interesting to note that this expansion contains precisely the same terms as the one derived by Haldane\cite{Haldane} for a boson annihilation 
operator in a Luttinger liquid.   We also see why the replacement of the exponential Jordan-Wigner string operator by a cosine form, 
its Hermitian part,  is not really necessary.  Keeping the complete expansion in Eq.  (\ref{bosexp}), we get the same set of operators 
either way. 

Now consider the effect of the term in Eq.  (\ref{mLR}) in the X-ray edge approach.  After the unitary transformation 
of Eq.  (\ref{transform}) the term in $S^-_j$ linear in the $\bar d$ operator, with momentum $k$, is:
\be S^-_j\propto \bar d e^{{-i\over \sqrt{2\pi K}}(\gamma_R\bar \varphi_R+\gamma_L\bar \varphi_L)}
\sum_{n\in Z}a_{n}'e^{i[(2n+1)k_F+\pi +k]j+i(2n+1)\sqrt{\pi K}\bar \phi}.\label{opsum}\ee
This expansion is similar to the one derived by Haldane\cite{Haldane} for a fermion annihilation operator.
The momentum $q$ at which the $n^{th}$ term contributes to the spectral function is:
\be q=(2n+1)k_F+\pi +k\ee
where the momentum $k$ must correspond to that of a high energy hole, $|k|<k_F$ in calculating $S^{+-}$ or to that 
of a high energy particle, 
$k_F<|k|<\pi$, in calculating $S^{-+}$. Note that by ignoring $m(x)$ in Sec. IIB we only considered the $n=0$ and $n=-1$ 
terms in the sum of Eq.  (\ref{opsum}). The $n^{th}$ term in the expansion of $S^-_j$ in Eq.  (\ref{opsum}) leads 
to a singular term in the transverse spectral function, 
\be S(q,\omega )\propto {1\over |\omega -\omega (q)|^\mu }\ee
at the energy $\omega (q)$ given by the energy of the corresponding particle or hole:
\be \omega_n(q)=\pm \epsilon [q-(2n+1)k_F-\pi ]. \label{omegan}\ee
We see that, for general $k_F$, these thresholds can occur at arbitrarily low energy. (For rational $k_F$ there is a 
finite number of them and there a lowest one at a non-zero energy.) The corresponding critical exponent is given by:
\be \mu^{(n)}=1-\nu_R^{(n)}-\nu_L^{(n)}\ee
where $\nu_{L,R}^{(n)}$ are the left and right scaling dimension of the $n^{th}$ operator in Eq.  (\ref{opsum}).
These obey:
\bea 
\label{eq:s-+1}
\nu_R^{(n)}+\nu_L^{(n)}&=&{1\over 4}\left[\left({\gamma_R\over \pi \sqrt{K}}+(2n+1)\sqrt{K}\right)^2
+\left({\gamma_L\over \pi \sqrt{K}}-(2n+1)\sqrt{K}\right)^2\right] \nonumber \\
&=&{1\over 2}\left[(2n+1)\sqrt{K}+{\gamma_R-\gamma_L\over 2\pi \sqrt{K}}\right]^2+{1 \over 2}\left({\gamma_L+\gamma_R \over 2\pi \sqrt{K}}\right)^2
.\eea
Thus the two largest values of $\mu$ occur for $n=0$ and $-1$ for $|\gamma_R-\gamma_L|/(2\pi K)<1$.
 At zero field, $\gamma_L=\gamma_R$ so this condition is satisfied. 
Also at small $\Delta$, $\gamma_{L,R}$ are $O(\Delta )$ so the condition is again satisfied. 
It should remain satisfied for a large range of field and $\Delta$ quite possibly including the 
entire Luttinger liquid regime, but without determining the $\gamma_{L,R}$ explicitly we can't 
determine this range. It certainly includes weak fields $h\ll J$ relevant to most ESR experiments. For zero 
field we have explicitly:
\be \nu_R^{(n)}+\nu_L^{(n)}={K\over 2}(2n+1)^2+{(1-K)^2\over 2K}\ee
so we see that the exponents for sub-dominant singularities ($n\neq 0$, $-1$) obey $\mu <-3/2$ for all 
$K$ in the Luttinger liquid regime $K>1/2$. At $\Delta =0$, 
\be \nu_R^{(n)}+\nu_L^{(n)}={1\over 2}(2n+1)^2\ee
and  $\mu <-7/2$ for all sub-dominant singularities. We expect that $\mu <0$ for all sub-dominant singularities 
a wide range of field and $\Delta$ including the weak field regime. So the approximation of 
dropping $m(x)$ made in sub-section IIB appears quite generally valid.

However, we must also consider the other terms in $m\propto \psi^\dagger_Rd$, $\psi^\dagger_Ld$. 
These give contributions to the transverse spectral functions proportional to matrix elements 
in the $\bar d$ space  containing more operators. Consider, for example, the case of a 
high energy  particle. Then, to first order in these operators we either obtain 
$\bar d(-\epsilon )\bar d^\dagger (0)|0>=0$ or else a matrix element:
\be <0|\bar d(x,t)\bar d(x-\epsilon ,t)\bar d^\dagger (-\epsilon )\bar d^\dagger (0)|0>.\label{dme}\ee
In fact, this is also zero as $\epsilon \to 0$ as follows from Wick's theorem and translation invariance:
\be <0|\bar d(x,t)\bar d(x-\epsilon ,t)\bar d^\dagger (-\epsilon )\bar d^\dagger (0)|0>=<0|\bar d(x,t)\bar d^\dagger (0,0)|0>^2
-<0|\bar d(x-\epsilon ,t)\bar d^\dagger (0,0)|0>^2\to 0.\ee
Now consider the higher order expansion in the terms in $m$ proportional to $\bar d$ and $\bar d^\dagger$. Even 
orders in the expansion give 
\be [\bar d(-\epsilon )\bar d^\dagger (-\epsilon )]^n\bar d^\dagger (0)|0>\propto \bar d^\dagger (0)|0>.\ee
The factor in the Green's function involving Fermi surface excitations has $S^-$ dressed by 
$n$  $\psi_{L,R}$ operators and $n$ $\psi^\dagger_{L,R}$ operators, all at the same point. The 
$(\psi_L^\dagger \psi_R)^n$ and $(\psi_R^\dagger \psi_L)^n$ terms just give contributions the 
same as Eq.  (\ref{opsum}), modifying the $a_n'$ coefficients. Other products give higher dimension operators 
using:
\be \psi^\dagger_L\psi_L\propto A + B\partial_x\phi_R+C\partial_x\phi_L\ee
et cetera. 
Odd terms in the expansion in the terms in $m$, which are proportional to $\bar d$ and $\bar d^\dagger$, give zero 
as $\epsilon \to 0$, since they are proportional to the same matrix element,  Eq.  (\ref{dme}). 

Till now we have considered the effect of particle excitation on $S^{-+}$. In general high-energy hole excitations, also could contribute to $S^{-+}$ as well as particle excitations to $S^{+-}$. Now we will find the effect of hole excitations on $S^{-+}$. Hole excitations only give us higher order corrections to $S^{-+}$, which have sub-dominant  singularities, but to complete our discussion we find the vanishing singularity exponent of hole excitation to $S^{-+}$; similar argument holds for particle excitation effects on $S^{+-}$.
In addition to the Eq. (\ref{first+-}), there is another term which contributes to $S^{-+}$ of the following form
\be \label{second+-} S^{-+}_j(t)=e^{i\pi j}e^{ik_f j}<\psi_R(j,t)\cos [\pi N_j(t)]\cos [\pi N_0(0)]\psi_R^{\dagger}(0,0)>.\ee 
by similar argument as discussed in section \ref{sec:xray} we could do the canonical and unitary transformations to decouple bosonic fields from high-energy excitations. Now the zero order term gives us the naive bosonization results; the interesting contribution comes from the first order expansion of $m\propto \psi^\dagger_R d_h$, $\psi^\dagger_L d_h$, where $d_h$ is hole creation operator.
The first order correction of expansion has the following form
\begin{eqnarray}
 S^{-+} & \propto  &<\bar{\psi}_R(x,t)\bar{\psi}_R(x-\epsilon,t)\bar{d}^{\dagger}_h(x-\epsilon,t)\cos [k_F x+\sqrt{\pi K} \bar{\phi}(x,t)]
e^{\frac{i}{\sqrt{2 \pi K}}[\gamma_R \bar{\varphi}_R(x,t)+\gamma_L\bar{ \varphi_L}(x,t)]}\nonumber \\
&\:& e^{\frac{-i}{\sqrt{2 \pi K}}[\gamma_R \bar{\varphi}_R(0,0)+\gamma_L\bar{ \varphi_L}(0,0)]}
\cos [\sqrt{\pi K} \bar{\phi}(0,0)]\bar{d}_h(\epsilon,0)\bar{\psi}_R^{\dagger}(\epsilon,0)\bar{\psi}_R^{\dagger}(0,0)>e^{i(2k_F-k_h+\pi)x} \nonumber 
\end{eqnarray}
Similar to Eq. (\ref{mom}) we could decompose $S^{-+}$ to two terms as follow
\be S^{-+}(x,t)=e^{i(\pi-k_h+k_F)x} I_h^-(x,t)+e^{i(\pi-k_h +3 k_F )x} I_h^+(x,t)\ee
where the expression for  $I_h^\pm$ is similar to Eq. (\ref{I}) with exponents given by
\bea
\bar{\nu}^{\pm}_R =  {1 \over 4}\left({\gamma_R \over \pi \sqrt{K}}+{2 \over \sqrt{K}} +(2 \pm 1) \sqrt{K}\right)^2 \nonumber \\
\bar{\nu}^{\pm}_L =  {1 \over 4}\left({\gamma_L \over \pi \sqrt{K}}+{2 \over \sqrt{K}} -(2 \pm 1) \sqrt{K}\right)^2 
\eea
In general, the higher order correction will include more powers of $\psi_R^{\dagger}\psi_L$ and general expression would be
\bea
\bar{\nu}^{(n)\pm}_R =  {1 \over 4}\left({\gamma_R \over \pi \sqrt{K}}+{2 \over \sqrt{K}} +(2n \pm 1) \sqrt{K}\right)^2 \nonumber \\
\bar{\nu}^{(n)\pm}_L =  {1 \over 4}\left({\gamma_L \over \pi \sqrt{K}}+{2 \over \sqrt{K}} -(2n \pm 1) \sqrt{K}\right)^2 
\eea
with momentum given by
\be 
k^{(n)} = \pi -k_h + (2n+1) k_F
\ee 
and the energy of the excitation for given momentum $q$ is
\be \bar{\omega}_n(q)=\pm \epsilon [-q+(2n+1)k_F+\pi ].\ee
$\omega_n$ and $\bar{\omega}_n$ for $n=1$, are depicted by dashed lines in Fig(\ref{fig:sing2}), which represents vanishing singularities of spectral function. The sum of exponents is
\bea
\label{eq:s-+2}
 \bar{\nu}_R^{(n)}+\bar{\nu}_L^{(n)}
&=&{1\over 2}\left[(2n+1)\sqrt{K}+{\gamma_R-\gamma_L\over 2\pi \sqrt{K}}\right]^2+{1 \over 2}\left({\gamma_L+\gamma_R \over 2\pi \sqrt{K}} + {2 \over \sqrt{K}}\right)^2
.\eea
We see that our results Eq. (\ref{eq:s-+1}) and (\ref{eq:s-+2}) are the same as Eq. (17) of Ref [\onlinecite{glazman09}], which is the singularity exponent for boson creation operator, upon 
identifying $\delta_{\pm}\to \gamma_{R/L}/\sqrt{K}$. The actual values of these phase shift parameters are in general different in the two models however, 
being determined by Galilean invariance in the Bose gas model. This correspondence might have been anticipated since a boson creation 
operator is related to the corresponding fermion one by a Jordan-Wigner string operator\cite{Haldane} just as is the $S^+_j$ operator.  Furthermore, 
the xxz model is equivalent to a lattice boson model with an infinite on-site repulsion which restricts the occupancy to $0$ or $1$. \\

Now we look at the effect of Umklapp scattering term at zero magnetic field. The Umklapp scattering term is  in the following form
\be \mathcal{H}_{U} = -g \left[e^{-4ik_F x} \psi^{\dagger}_R(x)\psi^{\dagger}_R(x) \psi_L(x) \psi_L(x) + \hbox{h.c}\right]\ee
At non-zero magnetic field we could ignore this term as it is highly oscillatory, due to  $e^{i4k_F x}$ prefactor. At zero field we have $4k_F=2\pi$; thus this term does not oscillate and we need a more careful treatment. In the bosonized form of the XXZ model, it can be seen that the Umklapp term is irrelevant for $0< \Delta <1$, and is marginal for $\Delta=1$, thus in this regime we could look at the effect of this term perturbatively. 
 We focus on the effect of this term on $S^{-+}$. In general the higher order Umklapp term could be written as follow
\be
U_{2\bar{m}+m}=\prod_{i=1}^{m+\bar{m}}(\psi^{\dagger}_R(z_i))^2 (\psi_L(z_i))^2 \prod_{j=1}^{\bar{m}}(\psi^{\dagger}_L(z'_j))^2 (\psi_R(z'_j))^2
\ee
Where $m$ and $\bar{m}$ are arbitrary integers, and this term is actually $2\bar{m}+m$ order in the Umklapp perturbation. These operators gives us zero corrections unless we keep higher powers of the $m(x)$ term, Eq. [\ref{mLR}], at least to power $2m$. Thus the most general non-zero term of both Umklapp and oscillatory term $m(x)$ is in the following form
\be P_{n,m,\bar{m}} = U_{2\bar{m}+m}(\psi^{\dagger}_L(x) \psi_R(x))^{m}(\psi^{\dagger}_L(0) \psi_R(0))^{m}\ee 
Now by bosonizing the above expression and plugging it into the definition of $S^{-+}$ we have
\bea
S^{-+} & \propto & \int <\bar{d}(x) \cos(k_F+\sqrt{\pi K}\varphi(x))e^{{-i \over \sqrt{2\pi K}}(\gamma_R \varphi_R(x) + \gamma_L \varphi_L(x))} e^{i\sqrt{4\pi K}(m-n)\varphi(x)} e^{-i2\sqrt{4\pi K}\sum^{m+\bar{m}}\varphi(z_j)} \nonumber \\
& \times & e^{i2\sqrt{4\pi K}\sum^{\bar{m}}\varphi(y_i)}\cos(\sqrt{\pi K} \varphi(0))e^{{i \over \sqrt{2\pi K}}(\gamma_R \varphi_R(0) + \gamma_L \varphi_L(0))}e^{i\sqrt{4\pi K}(n+m)\varphi(0)}\bar{d}^{\dagger}(0)> \prod^{\bar{m}} d^2y_i \prod^{m+\bar{m}} d^2z_j
\eea
The effect of this term on the singularity exponents of $S^{-+}$ could be evaluated by power counting and the result is given by the following expression
\bea
 \nu_R & = & \left({\gamma_R \over 2\pi \sqrt{K}}-n\sqrt{K} \pm {\sqrt{K}\over 2}\right)^2+K(m^2+2m+4\bar{m})-(m+2\bar{m}) \nonumber \\
  \nu_L & = & \left({\gamma_L \over 2\pi \sqrt{K}}+n\sqrt{K} \mp {\sqrt{K}\over 2}\right)^2+K(m^2+2m+4\bar{m})-(m+2\bar{m}) \nonumber \\
\eea
The $-(m+2\bar{m})$ term in above equations comes from the integration variables $z_j,y_i$. The overall exponent is given by the summation of these two exponents, thus we have
\be
\nu_R+\nu_L = {1 \over 2}\left({\gamma_R+\gamma_L \over 2\pi \sqrt{K}}\right)^2 +{1 \over 2 }\left({\gamma_L - \gamma_R \over 2\pi \sqrt{K}} + (2n\mp 1)\sqrt{K}\right)^2 +2K m^2 + 2(2\bar{m}+m)(2K-1) 
\ee
As we are considering the effect of Umklapp term at zero field, $4k_F = 2\pi$; these are singularity exponents of the spectral function at threshold frequencies given by Eq. [\ref{omegan}]. Compared to Eq. [\ref{eq:s-+1}], we see that for fixed $n$ higher order Umklapp terms give larger and larger exponents, and are more irrelevant. Therefore, at each threshold energy, we get a
set of singularities with progressively weaker exponents. In this case,
these corrections seem truly unimportant. That is, we don't get any new
singular energies, just sub-dominant corrections to the singularities at
the energies we already have. \\\

One important point is that if we take $n,m=0$ but arbitrary $\bar{m}$, at half filling and for $\Delta =1$ we have $2K-1 =0$; thus higher order non-chiral Umklapp corrections do not change the singularity exponents, based on power counting. But we should be careful at that limit, because our result was based on power counting; in general upon evaluating the integrals more carefully, there could be some logarithmic corrections to the correlations functions which could change the behaviour of spectral functions near singular frequencies.\\\

Till now we only considered the effect of Umklapp terms only at zero magnetic field. At finite magnetic field the Umklapp interaction has the following form
\be 
\mathcal{H}_U = g_U \cos 4(\sqrt{\pi K} \varphi(x) +(k_F-\pi/2a)x)
\ee
Where $a$ is the lattice spacing. In general this term is oscillatory and could be drop out at low energies. But for weak enough magnetic field, the wavelength of the oscillation is very long, therefore in that limit this term should be treated carefully. We claim that at low enough temperatures and weak magnetic field the Umklapp term affects neither the threshold frequency nor the singularity exponents, but it will change the overall behaviour of the spectral function; the reason is as follow. If we include the effect of Umklapp term perturbativley, it could be easily shown that such higher order terms can not change the oscillation wave-vector of the spectral functions. 
 It only modifies the non-oscillatory part of the spectral function without changing the oscillatory part.
 Therefore, if the oscillations wave-vector remains intact the threshold frequency does so.\\
 
The singularity exponents does not change because, to find the singularity exponents of spectral functions, we need to study the behaviour of spectral functions at frequencies, $\omega$, around the threshold frequencies, $\omega_L$. In principle the probe frequency could be chosen as close as possible to the singular frequency such that the Umklapp term effects would be irrelevant at those energy difference scales, $|\omega-\omega_L| \ll |(k_F-2\pi/a)v|$. Thus the Umklapp term would not change the singular exponent at $\omega_L$; we expect a cross over regime where the effect of Umklapp will be important at energies near to $|\omega-\omega_L| \approx (k_F-2\pi/a)v|$.
\section{ Electron spin resonance with Uniform Dzyaloshinskii-Moriya Interactions}

Electron spin resonance provides a sensitive probe of spin dynamics. A microwave field is weakly Zeeman coupled to the $q=0$ components of the spin operators. In the 
standard (Faraday) configuration, the microwave field is polarized perpendicular to a static magnetic field. For simplicity we restrict ourselves to the relatively 
simple situation of Eq.  (\ref{xxz}), with DM vector and magnetic field in the $z$-direction. 
Then, as discussed in Sec. I, a uniform DM interaction added to 
the xxz model of Eq.  (\ref{xxz}) simply shifts the parameters $J$ and $\Delta$ and the momentum, $q$ in the transverse spectral function.   Therefore the ESR adsorption intensity is proportional to the transverse spectral function at $q=\alpha = \arctan (D/J)$. Low temperature ESR measurements on quasi-1D antiferromagnets with uniform DM interactions 
could therefore probe the edge singularities predicted by X-ray edge methods that disagree with standard bosonization results due to the 
effects of band curvature. Such ESR results would be especially useful if they were done with circularly polarized microwave radiation since then 
$S^{-+}$ and $S^{+-}$ could be measured separately.  A further major challenge for such ESR experiments would be that the 
theoretical predictions give $S^{ss'}(\alpha ,\omega ,h)$ for fixed $h$ as a function of $\omega$. However, in an ESR experiment, $\omega$ is 
normally fixed at the resonant frequency of a microwave cavity and $h$ is varied. $\omega$ can only be varied by using a sequence of 
microwave cavities with different resonant frequencies. Alternatively, theoretical line-shapes could be produced for fixed $\omega$ 
and varying $h$ but these would be complicated since the critical exponent $\alpha$ varies with $h$.  For simplicity, 
we just discuss the line shape versus frequency at fixed $h$ here.  

We begin by discussing the $T=0$ limit. 
The ESR adsorption intensity can be simply read off from the results of Sec. II, III.  We first consider the case of  zero static field with the microwave field in 
the xy plane. Then the adsorption intensity has a lower threshold near which:
\be I(\omega )\propto {\theta (\omega -v\sin \alpha )\over (\omega -v\sin \alpha )^\mu }\ee
with $\mu = 2-1/(2K)-K$.  Here $v$ and $K$ are determined in terms of $\Delta_{eff}=\Delta \cos (\alpha )$ by Eq.  (\ref{vK}).  For small $\alpha$ 
we expect the results of standard bosonization to apply at somewhat higher energies, $\omega -v \alpha \gg  \alpha^3 J$. 
In this region we obtain:
\be I(\omega )\propto {(\omega +v \alpha )^{1+2\eta}\over (\omega -v\alpha )^{1-2\eta}}\label{Ist}\ee
with $\eta =(1-2K)^2/(8K)$. Note that $1-2\eta = 2-1/(4K)-K\neq \mu$, a different exponent than occurs at the threshold. As discussed in sub-section IIA,
$I(\omega )$ in Eq.  (\ref{Ist})  is non-monotonic, eventually passing through a minimum and starting to increase again as $\omega$ increases. However, since 
the formula is only valid in the low energy regime, $\omega \ll J$, whether or not this minimum occurs in the frequency region where the  
formula is valid depends on $\alpha$ and $\Delta_{eff}$. If $\Delta_{eff}\geq \cos \alpha$, the value resulting from $\Delta =1$, then 
the minimum predicted by standard bosonization is not in the region where the approximation is valid. In this case the intensity 
is monotone decreasing up to high frequencies where our techniques break down. $\Delta_{eff}$ is typically close to 1. In fact, with some 
assumptions about the higher energy levels of the magnetic ion, it is exactly one.\cite{kaplan} In this case, the transverse spectral function 
becomes the same as the longitudinal one discussed extensively in [\onlinecite{karbach, caux,rodrigo07, rodrigo08}]. The edge exponent 
has the value $\mu =1/2$, first obtained from the 2-spinon approximation
\cite{karbach} in this case and the standard bosonization prediction of Eq.  (\ref{Ist}), with $\eta =0$, reduces to a $\delta$-function which 
fails to capture many features of the actual spectral function for non-zero $\alpha$.  In particular there is a narrow peak of width  $\propto \alpha^3$ followed by a 
slowly decaying tail at higher $\omega$. 

At non-zero field, $h$, we may again use the results of Sec. II, III, which are less complete in this case.  
The X-ray edge results of Sec. III imply a quantum phase transition as the 
magnetic field is increased, occurring when the field-induced magnetization, $m(h)$, obeys $\alpha=H\equiv 2\pi m\approx Kh/(\pi /v)$ 
as can be seen from Fig.  [\ref{fig:sing2}]. $S^{-+}$ has a threshold singularity at a frequency $\omega_L \approx v(\alpha -H)$ for $H<\alpha \ll 1$.  A threshold 
singularity was also predicted in III for $S^{+-}$ in this field range, at a higher frequency, of approximately $v(\alpha +H)$. 
However, as discussed at the end of that sub-section, we expect this to be broadened.  On the other hand, for $H>\alpha$, 
we expect $S^{+-}$ to have a sharp 2-sided singularity at a frequency of approximately $v(H-\alpha )$.  The other threshold 
singularity predicted for $S^{+-}$ at $v(H+\alpha )$ in IIB is likely to be broadened.   (The presence of two peaks 
at these energies was first predicted in [\onlinecite{starykh}] and was observed experimentally in [\onlinecite{cscucle}].)  The precise energies and critical exponents 
for these singularities could be predicted by numerical Bethe ansatz calculations but analytic expressions are not 
available.  Assuming $\alpha $, $H\ll 1$, we expect the spectral 
functions to cross over to the form predicted by standard bosonization at energies somewhat higher than the singularities, 
$(H-\alpha )^2aJ\ll v|H-\alpha| \ll J$:
 \bea S^{-+}&\propto& { [\omega +v(H-\alpha )]^{1+2\eta}\over [\omega +v(\alpha -H)]^{1-2\eta}}\ \  (H<\alpha )\nonumber \\
 S^{+-}&\propto& { [\omega +v(\alpha -H)]^{1+2\eta}\over [\omega +v(H-\alpha )]^{1-2\eta}}\ \  (H>\alpha ).
 \eea
As discussed above and sketched in Fig.  [\ref{fig:transverse}], these functions are non-monotonic but the minimum only occurs 
in the energy region $\omega \ll J$, where the approximation holds, for a certain parameter range of $\alpha$, $\Delta$ and $h$. 
In addition to these dominant singularities, as discussed in IIIB,  we expect 
many additional weaker vanishing singularities extending down to low energies. 
  \subsubsection*{Finite Temperature Broadening}
  
At finite temperature the sharp peaks are broadened.  This can be calculated using standard bosonization for $(H-\alpha )^2aJ\ll T\ll J$. 
At finite $T$ the spectral function of Eq.  (\ref{Ist}) becomes:
\be I\propto \hbox{Im}\left[\sin (2\pi \eta )(2\pi T)^{\eta}B(1+\eta-i(\omega +v\alpha )/(4\pi T),-1-2\eta )B(\eta -i(\omega -v\alpha )/(4\pi T),1-2\eta )\right] .\ee
where $B(x,y)=\Gamma (x)\Gamma (y)/\Gamma (x+y)$ is the Euler beta function. For weak anisotropy, small $\alpha$ and $1-\Delta_{eff}$, 
and hence $\eta \ll 1$,  a Lorentzian line-shape occurs:
\be I\propto {1\over (\omega -v\alpha )^2+(4\pi T\eta )^2}\ee
with a similar broadening at finite $H$.  The width of the peak is $4\pi T\eta \approx 2T(1-\Delta_{eff})/\pi$ for $\Delta_{eff}$ close to 1. This is 
essentially the same result derived in [\onlinecite{Ian99}] for the ESR width due to exchange anisotropy parallel to the magnetic field.

\begin{figure}[htp]
\includegraphics[scale=1]{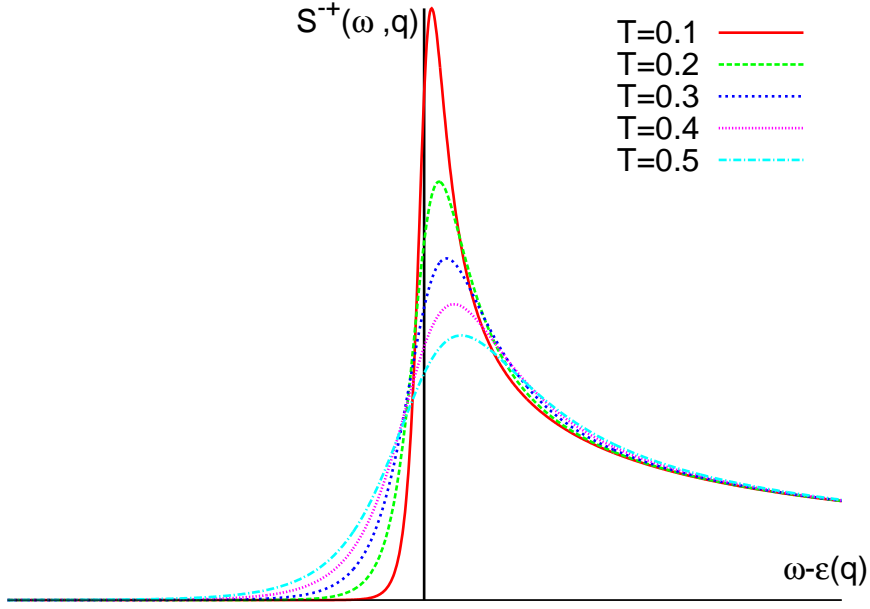}
\caption{The behaviour of $S^{+-}(\omega,q)$ spectral function for at finite temperature for the case $\nu_R=1/2$, $\nu_L=0$, corresponding 
to $\Delta =1$, $h=0$.  We see that for small enough T the broadening is asymmetric.}
\label{fig:finiteT}
\end{figure}

In section \ref{sec:xray} we found the singularity exponent and behaviour of the transverse spectral function near the thresholds at zero temperature using X-ray edge methods.
Now we look at the finite temperature effects on the spectral functions, and we find that finite temperature results in a non-Lorentzian broadening of the transverse 
spectral functions near the thresholds for  $|\omega-\epsilon(q)|,T\ll q^2/2m$.  
Non-zero temperature has two effects on the spectral function Eq. (\ref{S+-d}). First,  the Green's function for excitations near the fermi surface, $S^{(0)}$ is modified to the usual finite T form by a conformal transformation. Secondly, the Green's function for the $\bar d$
operators would be modified to finite T form - with step functions $\theta
(\epsilon )$ replaced by Fermi functions $n_F(\epsilon )$. This would allow a
hole contribution to $S^{-+}$ and a particle contribution to $S^{+-}$. 
However, at low $T\ll \epsilon(k_p)$ these would be negligible. So the only
important effect may be in $S^{(0)}$. Thus in this regime we have
\bea S^{-+}(q,\omega )&\propto& \int dx dt {e^{i[\omega -\epsilon (k_p^+)]t}\delta (x-ut) (2\pi T)^{\nu_R^+ +\nu_L^+}
\over (\sin(2\pi T(\epsilon + i(t-x/v))))^{\nu_R^+}(\sin(2\pi T(\epsilon +i(t+x/v))))^{\nu_L^+}}, \ \  (2k_F-\pi <q<\pi )   
\eea
By doing the integration over $x$ we have
\bea
\label{eq:xraynonzerot1}
 S^{-+}(q,\omega )&\propto& \int^{\infty}_{-\infty}  dt {e^{i[\omega -\epsilon (k_p^+)]t} (2\pi T)^{\nu_R^+ +\nu_L^+}
\over (\sin(2\pi T(\epsilon +i (1-u/v)t)))^{\nu_R^+}(\sin(2\pi T(\epsilon +i (1+u/v)t)))^{\nu_L^+}}, \ \  (2k_F-\pi <q<\pi )  
\eea
By evaluating the above integral we can get the finite $T$ behaviour of the spectral function. From the above equation it can readily be seen that the spectral function is a pure real number;
 in the appendix we will prove that it is positive too. 
Let us look at the behaviour of above spectral function for  $2k_F-\pi <q<\pi$, at $h=0$ and $\Delta_{eff}\approx 1$, small anisotropy $\eta\ll1$ so that 
$\nu_R=1/2-\eta$ and $\nu_L=0$. In this regime we have
\be
\label{eq:xraynonzerot2}
S^{-+} \propto \frac{1}{T^{1-\nu_R}}Re \left[e^{-i\pi \nu_R/2 }B\left(-i\frac{\omega-\epsilon (k_p^+)}{4\pi T(u-v)}+\frac{\nu_R}{2},1-\nu_R \right)\right] 
\ee
In Fig. [\ref{fig:finiteT}] we have depicted, the spectral function for different values of temperature. As temperature gets higher the broadening increases; as  can be seen, for small $T$ 
the broadening is asymmetric and so is non-Lorentzian. Furthermore, the width is $O(T)$, not suppressed by a factor of $1-\Delta_{eff}$ as predicted by 
standard bosonization. 

Since the effects predicted by the new theoretical methods occur at very low energy scales, a highly one-dimensional 
spin compound would probably be needed to observe them, in order that three dimensional exchange processes would be 
negligible. Furthermore, materials like KCuGaF$_6$ with both uniform and staggered intra-chain DM interactions may 
not be suitable since the staggered DM interactions tend to have a larger effect than the uniform ones.\cite{Ian99}
Thus identifying the right material to test these predictions remains an open challenge. An alternative approach might 
be to study metallic quantum wires with spin-orbit couplings. It was shown in [\onlinecite{starykh}] that 
analogous phenomena occur in that system.

\section{Conclusions}
By applying X-ray edge techniques, we have obtained results on the transverse spectral function of the xxz spin chain in 
a magnetic field. We illustrated why standard bosonization techniques fail near threshold energies, 
even when these occur at low energy. In the zero field case we have exactly determined a critical exponent governing the lower edge 
singularity, for all $|\Delta |<1$ and all wave-vector. For the finite field case, we have shown how this exponent can be determined 
from parameters which can be obtained from solving Bethe ansatz equations and which also determine the behaviour
of the longitudinal structure function, fermion spectral function and the finite size spectrum. We have argued that, for general magnetization, 
a large number of increasingly weaker singularities occur in the spectral function, extending all the way 
down to zero energy. We derived results for the finite temperature spectral function using X-ray edge methods, 
obtaining strikingly different behaviour than that given by standard bosonization at $0<1-\Delta_{eff}\ll 1$. 
The line-shape is non-Lorentzian and the line width is O($T$), unsuppressed by $1-\Delta_{eff}$. We pointed 
out that electron spin resonance measurements on spin chain compounds with uniform Dzyaloshinskii-Moriya 
interactions would provide a way of experimentally confirming, for the first time, the new bosonization results 
being obtained on spin chains, using X-ray edge techniques. 

In the case of staggered DM interactions, the most interesting 
ESR signal occurs when the magnetic field is transverse to the DM vector. This may also be the case 
for uniform DM interactions, but we leave this for future work.

We thank J-S Caux, Ion Garate, Leonid Glazman, Rodrigo Pereira, Oleg Starykh, Hidekazu Tanaka and Izumi Umegaki for helpful discussions. This 
research was supported by NSERC and CIfAR.

\appendix
\section{Positivity of spectral function }

In this appendix we will prove that spectral function Eq. (\ref{eq:xraynonzerot1}) is positive. We have
\bea
\label{eq:xraynonzerot22}
 S^{-+}(q,\omega )&\propto& \int^{\infty}_{-\infty}  dt {e^{i[\omega -\epsilon (k_p)]t} (2\pi T)^{\nu_R +\nu_L}
\over (\sin(2\pi T(\epsilon +i (1-u/v)t)))^{\nu_R}(\sin(2\pi T(\epsilon +i (1+u/v) t)))^{\nu_L}}  \nonumber \\
&\propto& \left[ \int^{\infty}_{0}+ \int^{0}_{-\infty}\right]  dt {e^{i[\omega -\epsilon (k_p)]t} (2\pi T)^{\nu_R +\nu_L}
\over (\sin(2\pi T(\epsilon +i (1-u/v) t)))^{\nu_R}(\sin(2\pi T(\epsilon +i (1+u/v) t)))^{\nu_L}}
\eea
By defining the branch cut on $(-i \infty,-i\epsilon)\bigcup (i\epsilon ,i\infty)$ and doing the change of variable $t \rightarrow -t$ in second integral, we have
\bea
S^{-+}(q,\omega )&\propto& Re\left[\exp\left(-i \frac{\pi}{2} (\nu_L+\hbox{sgn}(1-u/v)\nu_R)\right)\int^{\infty}_{0} dt {(2\pi T)^{\nu_R +\nu_L-1} \: e^{i[\omega -\epsilon (k_p)]t/(2\pi T)} 
\over |\sinh((1-u/v) t)|^{\nu_R}|\sinh((1+u/v) t)|^{\nu_L}}
\right]
\eea
We will prove shortly that above integral is positive in general, 
but let us first look at the special cases of either $\nu_R=0$ or $\nu_L=0$, which is relevant for weak anisotropy at zero magnetic field, so $u<v$.\\
Suppose that $\nu_L=0$ thus we have
\bea
S^{-+}(q,\omega )&\propto& (2\pi T)^{\nu_R-1}Re\left[\exp\left(-i \frac{\pi}{2}\nu_R\right)\int^{\infty}_{0} dt {e^{i[\omega -\epsilon (k_p)]t/(2\pi T)} 
\over |\sinh((1-u/v) t)|^{\nu_R}} \right] \nonumber \\
&\propto & (2\pi T)^{\nu_R-1} Re\left(\exp\left(-i \frac{\pi}{2}\nu_R\right) B\left[-i \frac{\omega -\epsilon (k_p)}{4\pi T(1-u/v)}+\nu_R/2,1-\nu_R\right]\right)
\eea
Where $B[x,y]$ is Euler's beta function. Let's define $W \equiv \frac{\omega -\epsilon (k_p)}{4\pi T(1-u/v)}$ , and by using the definition of beta function in terms of gamma function we have
\be
S^{-+}(q,\omega )\propto (2\pi T)^{\nu_R-1}  Re\left(\exp\left(-i \frac{\pi}{2}\nu_R\right)\frac{\Gamma[-iW +\nu_R/2]\Gamma[1-\nu_R]}{\Gamma[1-(iW+\nu_R/2)]}\right)
\ee 
 Then by using the identity that $\Gamma[1-z]\Gamma[z]=\pi/\sin(\pi z)$ we have
\be
S^{-+}(q,\omega )\propto (2\pi T)^{\nu_R-1} \Gamma[-iW +\nu_R/2]\Gamma[iW +\nu_R/2]\Gamma[1-\nu_R] Re\left(\exp\left(- i \frac{\pi}{2}\nu_R\right)\sin\pi(iW+\nu_R/2)\right)
\ee 
by writing the sine function in exponential form, we finally get
\be
S^{-+}(q,\omega )\propto (2\pi T)^{\nu_R-1} \left|\Gamma[iW +\nu_R/2] \right|^2 \Gamma[1-\nu_R] e^{\pi W}\sin(\pi \nu_R)
\ee 
so we see the spectral weight is positive, for all values of  $\nu_R$. This expression also holds for the case of $u>v$.\\
Actually what we have shown is that the Fourier transform of $1/\sin(2\pi T (\epsilon +i (1-u/v)t))^\nu$ is given by a real positive function, in the following form
\be
 F.T\left[{1 \over \sin(2\pi T (\epsilon +i (1-u/v)t))^\nu}\right]=(2\pi T)^{\nu-1} \left|\Gamma\left[\nu/2+i{\omega \over 4\pi T(1-u/v)} \right]\right|^2\Gamma[1-\nu] e^{\pi \omega/(4\pi T(1-u/v))}\sin(\pi \nu)
\ee
Where $F.T$ stands for Fourier Transform. Now by using above equation and taking the convolution  of Eq. (\ref{eq:xraynonzerot2}), we have
\bea
S^{-+}(q,\omega )&\propto&(2\pi T)^{\nu_R +\nu_L-2}  \Gamma[1-\nu_R] \: \sin(\pi \nu_R) \: \Gamma[1-\nu_L] \: \sin(\pi \nu_L) \: \int^{\infty}_{-\infty}  \: d\omega_R \: d\omega_L  \: \delta(\omega -\epsilon (k_p)-\omega_R-\omega_L)  \nonumber \\
&\times & \left|\Gamma\left[\nu_R/2+i{\omega_R \over 4\pi T(1-u/v)} \right]\times \Gamma\left[\nu_L/2+i{\omega_L \over 4\pi T(1+u/v)} \right]\right|^2 e^{\pi \omega_R/(4\pi T(1-u/v))} e^{\pi \omega_L/(4\pi T(1+u/v))}
\eea
We see that all the functions in above equation are positive; thus whole the integral is positive. 

\begin{thebibliography}{50}
\bibitem{AG} In Ref. (\onlinecite{glazman09}) it was stated that the transverse spectral function exponent had been derived by X-ray edge methods for 
$h=0$, $\Delta =1$ and that it agreed with the result of Ref. (\onlinecite{rodrigo08}) for the longitudinal spectral function. However, no details 
of the derivation 
were given nor was the exponent  given for any other values of $\Delta $ or $h$.
\bibitem{ph} In [\onlinecite{rodrigo09}] the case $k_F<\pi /2$ was considered. To apply those results to the case $k_F>\pi /2$, 
that we are considering here, we must make a particle-hole transformation.
\end{thebibliography}



\end{document}